\documentclass[12pt]{article}

\renewcommand{\baselinestretch}{2.0}

\def\be{\begin{equation}}
\def\ee{\end{equation}}
\def\bea{\begin{eqnarray}}
\def\eea{\end{eqnarray}}

\def\ol{\overline}

\def\nnmb{\nonumber}

\def\mbmm{{\rm \makebox[2em]{ }}}

\def\1{\={1}}
\def\2{\={2}}
\def\AA{$\mid$}
\def\MM{$_{\:}\rangle\mid$}
\def\NA{$\mid\! 0_{1}\rangle$}
\def\NB{$\mid\! 0_{2}\rangle$}
\def\NC{$\mid\! 0_{3}\rangle$}
\def\ND{$\mid\! 0_{4}\rangle$}
\def\NE{$\mid\! 0_{5}\rangle$}
\def\NF{$\mid\! 0_{6}\rangle$}

\def\ZZ{$\rangle$}
\def\ZA{\ZZ & \AA}

\def\BG{$\mid\,$}         % G for Gap after |

\def\p1{ 1$\;$}
\def\m1{-1$\;$}
\def\pB{ 2$\;$}
\def\mB{-2$\;$}
\def\pD{ 4$\;$}
\def\mD{-4$\;$}
\def\mwo{$\;$-1$\;$}
\def\mqrt{\!\!\!\!$-$\sqrt}

\begin{document}

\renewcommand{\thepage}{\arabic{page}}
\setcounter{page}{1}

\makebox{ }  \hfill NUHEP-TH-00-79 \\ 
%\makebox{ }  

\begin{center}

{\large \bf 
$E_6$ unification model building II.\\ 
Clebsch-Gordan coefficients of 78$\otimes$78 
}

\vskip 0.4in

Gregory W.~Anderson 
and
Tom\'{a}\v{s} Bla\v{z}ek$^{*}$ 
                   %$^{a*}$ 

\vskip 0.2in

%%$^a$
    {\em Department of Physics and Astronomy, Northwestern University,}\\
    {\em 2145 Sheridan Road, Evanston, IL 60208, USA}\\
\vskip 0.05in

e-mail: {\em anderson@susy.phys.nwu.edu, blazek@heppc19.phys.nwu.edu}

\vskip 0.2in

\end{center}

\vskip 0.1in

%********1*********2*********3*********4*********5*********6*********7*********8
\begin{abstract}

We have computed the Clebsch-Gordan coefficients
for the product (000001) $\otimes$ (000001), where
(000001) is the adjoint 78-dimensional representation
of $E_6$.
The results are presented for the dominant weights of the
irreducible representations in this product.
As a simple application we express the singlet operator in
${\bf 27}\otimes{\bf 78}\otimes{\bf \ol{27}}$ in terms of multiplets
of the Standard Model gauge group.

\end{abstract}

\vskip 1.5in

%PACS numbers: 02.20.Hj, 11.30.Ly 

\vskip 0.1in

$^*${\footnotesize On leave of absence from 
the Dept. of Theoretical Physics, Comenius Univ., Bratislava, Slovakia.}

\renewcommand{\baselinestretch}{1.1}

\renewcommand{\thesection}{\Roman{section}}
\setcounter{section}{0}

\renewcommand{\thetable}{\Roman{table}}
\setcounter{table}{0}

\section{Introduction}

The group $E_6$ is a promising and popular candidate for a grand unified group.
Despite the fact that it has received consideration for
over twenty years \cite{grs}, $E_6$ model building has not been extensively 
developed due to mathematical complexities associated with a rank 6 
exceptional Lie group. 
The Clebsch-Gordan coefficients (CGCs), for instance, have only 
been known for the products of two fundamental irreducible 
representations (irreps) of the lowest dimensionality:
{\bf 27} or ${\bf \ol{27}}$ \cite{pat_E6,abI}. To our knowledge, the 
CGCs for higher dimensional irreps of $E_6$ have never been computed. 
The difficulties are not related just to a large number of independent 
states in the weight system, but also to the construction of bases for 
states with degenerate weights. The latter problem is trivial
for smaller groups like {\em e.g.} $SU(2)$ or $SU(3)$
which are of the highest interest for elementary particle phenomenology,
and can be avoided altogether by the use of tensor methods and Young tableaux.
However, for $E_6$ it becomes a progressively larger obstacle for higher dimensional
irreps. In the 27-dimensional irreps of $E_6$, 
the basis is simply the weight system due to the fact that 
each weight state in the ${\bf 27}$  and ${\bf \ol{27}}$ 
is non-degenerate. The irreps with dimensionality
78, 351, and 650 are slightly more complicated, but do not pose a serious technical
challenge, since the bases may be chosen to
coincide with the weight system 
obtained by the application of ladder operators (group generators 
outside the Cartan subalgebra) despite the presence of degenerate weights. 
For the larger irreps, when derived by a method of successive 
lowerings from the highest weight, one obtains weight subspaces 
with the number of vectors by far exceeding the dimensionality 
of the weight subspace. As a randomly selected example, 
by constructing a complete set of states 
we found that the (-1,1,-1,1,-1,1) weight subspace
of {\bf 2430} $\subset$ {\bf 78}$\,\otimes\,${\bf 78} 
contained 28 unique vectors which span an 11-dimensional subspace. 
For the (0,0,0,0,0,0) weight subspace, our analysis resulted in
185 distinct linearly dependent states while dimensionality of the subspace is 36.

Several methods have been suggested which could be used to address this problem.
One could proceed by methods based on group 
subalgebras \cite{subalgbr} which, however, become laborious for
a rank 6 group. A more elegant method has been proposed in the analysis 
of Li {\it et al.} \cite{pat_Verma}, which introduces a 
set of rules for the construction of bases in irreducible
representation spaces of simple Lie algebras based on the unpublished ideas 
of D.N.~Verma. The bases are specified
in terms of sequences of lowering operators applied 
to the highest weight of the representation. The ordering is derived from
the opposite involution: a sequence of Weyl reflections which transforms
every positive root into a negative root. While the opposite involution is not 
unique, the exponents of the lowering operators in the involution satisfy 
basis-defining inequalities which are unique for a specific involution, 
if they exist.
\footnote  
{
The simplest example for such an inequality would be the $SU(2)$ 
relation $2|m| \leq 2\ell$, understood in the sense that the $SU(2)$ ladder
operator can be applied to the highest weight ($2\ell$) up to $2|m|$ times 
when constructing a particular representation. 
}
The same study, however, finds it difficult to apply the method to 
exceptional groups $E_6$  and $F_4$.\cite{pat_Verma}
The basis-defining inequalities for these two Lie groups are unknown 
while these inequalities are provided for
all other simple Lie groups with rank $n\leq 6$.
In light of these studies, our approach is rather pragmatic:
we adopt a straightforward procedure which probes all possible lowerings and calculates 
the complete set of states in the product, starting from the highest weight state
of the highest irreducible representation. While the method is straightforward,
due to a large number of degenerate weights and non-trivial lowering rules 
the task is technically quite complex. 
The closest similar computation to our knowledge has only been done for 
the product of two adjoints in $SU(5)$. \cite{pat_SU5}

The purpose of this paper is to present the results of our computation 
of the CGCs for the product of two adjoint representations in $E_6$.
These results are useful and necessary tools for building complete models
based on the unified group $E_6$. 
The paper follows our earlier work \cite{abI} where the
CGCs for the ${\bf 27}\,\otimes\,{\bf \ol{27}}$ were calculated
and the embeddings of the Standard Model fields into the {\bf 27},
the fundamental representation of $E_6$ have been listed.
In section 2, we present some basic theoretical background for the
computation. Section 3 contains our results for the dominant
weights in {\bf 78}$\otimes${\bf 78}. In section 4, we conclude with an application 
which shows how the singlet piece of ${\bf 27}\otimes{\bf 78}\otimes{\bf \ol{27}}$ 
can be expressed in terms of multiplets of the Standard Model gauge group.

\section{Theoretical Background}

We seek the construction of the CGCs in the $E_6$ tensor product
\be
 {\bf 78} \otimes {\bf 78} = {\bf 2430} \oplus {\bf 2925} 
                      \oplus {\bf 650}  \oplus {\bf 78} \oplus {\bf 1},
\ee
or, equivalently, in terms of the highest weights of each irrep
\be
 (000001) \otimes (000001) = (000002) \oplus (001000)
                      \oplus (100010) \oplus (000001) \oplus (000000).
\label{eq:hwss}
\ee

Our conventions for the root system of $E_6$ and other notation follow  
refs. \cite{Slansky_PhysRep}, \cite{pat_E6}, and \cite{abI}.
The group algebra includes 
\be
[H_i,H_j] = 0, \;\;\;
[H_i,E_{\alpha_j}] = (\alpha_j)_{_i} \, E_{\alpha_j}, \;\;\;
[E_{\alpha_j},E_{-\alpha_j}] = H_j
\label{eq:algbr}
\ee
(no implicit sum over repeating indices). The generators $H$ form the Cartan
subalgebra. The generators $E$ are the ladder operators and correspond to 
non-zero roots.
For simple roots $(\alpha_j)_{_i} = (\alpha_i,\alpha_j) = A_{ij}$, where
$A$ is the Cartan matrix
\be
     A = \left( \begin{array}{rrrrrr}
                               2&-1& 0& 0& 0& 0\\ 
                              -1& 2&-1& 0& 0& 0\\
                               0&-1& 2&-1& 0&-1\\
                               0& 0&-1& 2&-1& 0\\
                               0& 0& 0&-1& 2& 0\\
                               0& 0&-1& 0& 0& 2
                \end{array} \right).
\label{eq:Aij}
\ee

The weight system of the {\bf 78} coincides with the root system and we can set
$H_i|\alpha_j\rangle = (\alpha_j)_i\,|\alpha_j\rangle$.
The normalization of the generators satisfies
\footnote
{
In our conventions, all states are normalized to 1.
} 
\be
Tr(H_iH_j) = A_{ij}\,Tr(E_{\alpha_j}E_{-\alpha_j}).
\label{eq:normHH}
\ee
This is consistent with the algebra, eq.(\ref{eq:algbr}), and the 
lowering rules discussed below.

The lowering rules for the {\bf 78} are derived from the
lowering rules for the fundamental representations and the Clebsch-Gordan 
decomposition of the {\bf 78} states into the product of the {\bf 27}
and ${\bf \ol{27}}$ states \cite{abI}. This is especially important
for the six-fold degenerate zero weight of the {\bf 78}. 
The corresponding states can form an orthogonal basis 
$|\tilde{0}_i\rangle,\:(i=1,..,6)$, as is assumed in (\ref{eq:normHH}) or,
alternatively, one can consider a non-orthogonal basis $|0_i\rangle$
with each state specified by the last lowering: 
$|0_i\rangle\: \propto E_{-\alpha_i}|\alpha_i\rangle$, where 
$\alpha_i$ is a simple root. Based on the results of \cite{abI},
the inner product of the two basis states is in this case
\be
   \langle 0_i|\, 0_j \rangle\: = \frac{1}{2}\,A^0_{ij}, 
\label{eq:0i0j}
\ee
with $A^0_{ij}=|A_{ij}|$.
There is a non-singular transformation between the two bases,
\be
|\tilde{0}_i\rangle\: = \sum_{j=1}^6 C_{ij} \, |0_j\rangle,
\label{eq:Cdef}
\ee
which is non-unitary and corresponds to the projections
of simple roots onto an orthogonal basis.
Clearly, one is free to choose many different orthogonal bases
and a particular selection in the grand unified model building
will depend on the way how the $E_6$ symmetry is broken. For any choice,
however, $\,C^T C = 2 (A^0)^{-1} \equiv 2 G^0$. 
We find
\be
     G^0 = \frac{1}{3} 
           \left( \begin{array}{rrrrrr}
                               4& -5&  6& -4&  2& -3\\ 
                              -5& 10&-12&  8& -4&  6\\
                               6&-12& 18&-12&  6& -9\\
                              -4&  8&-12& 10& -5&  6\\
                               2& -4&  6& -5&  4& -3\\
                              -3&  6& -9&  6& -3&  6
                \end{array} \right)
\label{eq:G0ij}
\ee
and that is, up to signs, the weight space metric $G$ of $E_6$ \cite{Slansky_PhysRep}. 
Note that 
\be
    \sum_{k=1}^6 |\tilde{0}_k\rangle\,|\tilde{0}_k\rangle = 
    \sum_{i,j=1}^6     |0_i\rangle\: 2 G^0_{ij}\: |0_j\rangle.
\label{eq:00}
\ee
As a particular example of the C matrix consider
\be
     C =  \left( \begin{array}{rrrrrr}
      1 & -2 & 2 & -1 & 0 & -1 \\ 
    -\frac{1}{\sqrt{3}} & \frac{2}{\sqrt{3}} &  
    -\frac{4}{\sqrt{3}} & \frac{3}{\sqrt{3}} &  
    -\frac{2}{\sqrt{3}} & \frac{3}{\sqrt{3}}  \\
      1 & -1 & 1 & -1 & 1 &  0 \\ 
                                    \frac{1}{\sqrt{15}} & \frac{1}{\sqrt{15}} &  
                                    \frac{1}{\sqrt{15}} & \frac{3}{\sqrt{15}} &  
                                   -\frac{1}{\sqrt{15}} &   0 \\
    -\frac{1}{\sqrt{10}} &-\frac{1}{\sqrt{10}} &  
     \frac{4}{\sqrt{10}} &-\frac{3}{\sqrt{10}} &  
     \frac{1}{\sqrt{10}} &   0 \\
                                    \frac{1}{\sqrt{6}}  & \frac{1}{\sqrt{6}} &  
                                          0             &-\frac{1}{\sqrt{6}} &  
                                    -\frac{1}{\sqrt{6}} &   0 
                \end{array} \right).
\label{eq:Cij}
\ee
The first three rows of this matrix correspond to the projections 
onto the zero roots of the Standard Model gauge groups 
(two roots of the $SU(3)$ and one root of the $SU(2)$, respectively).
$|\tilde{0}_4\rangle$ lies in the hypercharge direction, and together
with the first three completes the zero weight space of the $SU(5)$ subgroup
of $E_6$.
In the same way, $|\tilde{0}_5\rangle$ lies in the direction of the $U(1)$
which is contained in the $SO(10)$ (it was called $U(1)_r$ in \cite{abI}),
and $|\tilde{0}_6\rangle$ in the direction of the $U(1)_t$, which is
perpendicular to the $SO(10)$ subgroup. Obviously, other branching chains
of $E_6$ would result in a modified $C$ matrix.

Next we specify the lowering rules. A {\bf 78} weight
state $|w\rangle$ of weight $(w)=(w_1,\ldots w_6)$ 
is lowered by 
$E_{-\alpha_i}$, with $\alpha_i$ being a simple root, according to
\be
   E_{-\alpha_i}|\, w\rangle\, = N_{-\alpha_i,w}|\, w-\alpha_i\rangle,
\label{eq:dfnN}
\ee 
which means that the new weight is always equal to $(w-\alpha_i)$ 
provided the new state exists. It is assumed that $N_{-\alpha_i,w}=0$
if the new state does not exist. For a non-zero weight and weight not 
equal to a simple root, the new state exists if the respective weight
(Dynkin) coordinate $w_i > 0$. In this case, $N_{-\alpha_i,w}=+1$.
For weight $(w)$ whose coordinates coincide with the coordinates of
simple root $\alpha_j$ the new state exists only 
if $j=i$, and $N_{-\alpha_i,\alpha_i}=+\sqrt{2}$. 
Finally for a zero weight, $(w) = (0)_j$, 
the new state exists if $A_{ij}\neq 0$, and 
$N_{-\alpha_i,0_j}=+|A_{ij}|/\sqrt{2}$. 
All lowering rules are accounted for in a single relation
\be
N_{-\alpha_i,(w)_j} = + [\, w_i\, +\, |N_{-\alpha_i,w+\alpha_i}|^2\;\; 
                      | \langle (w)_j\, |\, (w)_i\rangle|^2\;\; ]^{1/2}.
\label{eq:N-}
\ee
which relates the lowerings among the adjacent levels.
In this relation, the subscript on $(w)$ is only relevant for the degenerate
zero weights and can be ignored for all other weights of the {\bf 78}. 
Note that the zero weight states which we use to derive the 
CGCs in the {\bf 78}$\otimes${\bf 78} tensor product belong to the
non-orthogonal basis discussed above. In this case, the inner product entering
(\ref{eq:N-}) is given by eq.(\ref{eq:0i0j}). Relation (\ref{eq:N-} )
can be easily derived, up to the sign convention, from the group algebra, 
eq.(\ref{eq:algbr}),
using the property $E_{-\alpha_i}=E_{\alpha_i}^\dagger$.
The recursive relation (\ref{eq:N-}) must be further generalized
for weight systems with degenerate weights on successive levels~\cite{abIII}. 

\section{Clebsch-Gordan coefficients for ${\bf 78}\otimes {\bf 78}$}

As discussed in \cite{pat_E6}, it is sufficient to present the tensor
decomposition of the dominant weight states in the product. 
The CGCs of the other states can be obtained with the help of the 
charged conjugate operators introduced in \cite{pat_R} (or by direct lowerings).
Examples of applications of these operators can be found in \cite{abI}. 

The dominant weights in the product ${\bf 78}\otimes {\bf 78}$
are listed on the right side of eq.(\ref{eq:hwss}). 
We start with the highest weight state (or level 0) of the 2430-dimensional 
$(000002)$ irrep:
\be
 \mid\!\! 000002\rangle  = \,\mid\!\! 000001\rangle \mid\!\! 000001\rangle.
\ee
The first lowering leads to $(001000)$ which is another dominant weight. 
Following the rules outlined in the previous section the level 1
state of the $(000002)$ irrep consists of the symmetric combination 
\be
 \mid\!\! 001000\rangle  = \frac{1}{\sqrt{2}}\,(
                         \,\mid\!\! 000001\rangle \mid\!\! 00100\bar{1}\rangle 
                        +\,\mid\!\! 00100\bar{1}\rangle \mid\!\! 000001\rangle\:),
\label{eq:001000s}
\ee
while the orthogonal antisymmetric combination 
\be
 \mid\!\! 001000\rangle  = \frac{1}{\sqrt{2}}\,(
                         \,\mid\!\! 000001\rangle \mid\!\! 00100\bar{1}\rangle 
                        -\,\mid\!\! 00100\bar{1}\rangle \mid\!\! 000001\rangle)
\ee
represents the highest weight state of the 2925-dimensional $(001000)$ irrep.
($\bar{x}\equiv -x$ is used throughout this paper.) 

The next dominant weight, $(100010)$, is reached at level 6 of the {\bf 2430}
and is 3-fold degenerate (level 5 and 4-fold degenerate, in case of 
the {\bf 2925}). The state orthogonal to both of these irreducible subspaces 
is symmetric and becomes the
highest weight state of the 650-dimensional irrep. The $(000001)$ dominant
weight is then obtained at level 11 (10, 5) of the {\bf 2430}
({\bf 2925}, {\bf 650}) and is 11-fold (15-fold, 5-fold) degenerate.
The reducible $(000001)$ weight subspace is, however, 32-dimensional. The
extra orthogonal state 
is antisymmetric and represents the highest weight state of the {\bf 78}.
Finally, at level 22 (21, 16, 11) of the {\bf 2430}
({\bf 2925}, {\bf 650}, {\bf 78}) we get the 36-fold (45-fold, 20-fold, 6-fold) 
degenerate $(000000)$ weight. This reducible subspace is 108-dimensional
and leaves room for one singlet state, which is symmetric.  

For each dominant weight subspace at level $n$ the basis states are specified 
by their respective lowering paths: 
sequences of integers $i_n\ldots i_2i_1,\; 1\!\le\! i_k\!\le\! 6$. 
This is a shorthand notation for the sequence of lowering operators 
$E_{-\alpha_{i_n}}\ldots E_{-\alpha_{i_2}}\,E_{-\alpha_{i_1}}$ which is to be 
applied from right to left to the highest weight state 
in order to obtain a basis state. 
Lowering paths for the basis states of the {\bf 2430} and {\bf 2925}
are presented in tables \ref{t:paths_2430} and \ref{t:paths_2925}. 
Lowering paths relevant for the remaining irreps can be found 
in table 1 in \cite{abI}. Lowering paths are in general not unique.
This is not of much concern for the {\bf 650} or {\bf 78} since 
following different paths always yields the same basis states for these two irreps. 
However, this convenient property is no longer true for the {\bf 2430} and {\bf 2925}
where the number of distinct, albeit linearly dependent states may by far exceed the dimensionality of
the weight subspace.~\footnote 
{
Some examples from our numerical procedure were given in the Introduction.
}

Tables \ref{t:cgc100010}-\ref{t:cgc000000_2925_650_78_S} contain the Clebsch-Gordan coefficients for
the dominant weight states in {\bf 78}$\otimes${\bf 78}. 
In tables \ref{t:cgc000000_2430}-\ref{t:cgc000000_2925_650_78_S}, after showing the CGCs for 
a combination of states \AA$x$\ZZ\AA$\bar{x}$\ZZ\ we no longer show the CGCs for \AA$\bar{x}$\ZZ\AA$x$\ZZ.
The latter is either the same as the former for the symmetric $(000002)$, $(100010)$, and $(000000)$ 
irreps, or opposite in sign for the antisymmetric $(001000)$ and $(000001)$ irreps. 
For brevity $A$ in table \ref{t:cgc000001} stands for the adjoint $(000001)$ irrep,
and similarly, $S$ in table \ref{t:cgc000000_2925_650_78_S} denotes
the singlet.
Numbering of the degenerate
states is consistent with tables \ref{t:paths_2430} and \ref{t:paths_2925}, and 
table 1 in \cite{abI}.

The decomposition of the singlet (last column of table \ref{t:cgc000000_2925_650_78_S})
takes a very simple form:
\be
          S \equiv \: \mid\!\! 000000\rangle = 
         \frac{1}{\sqrt{78}}\: [\:2G^0_{ij}  \mid\! 0_i\rangle\!\mid\! 0_j\rangle 
                                         + \sum_{k=1}^{72} (-1)^{\ell+1}  \,\mid\! x_k\rangle\!\mid\! \bar{x}_k\rangle\,],
\ee
where $k$ enumerates the non-degenerate weight states of the $(000001)$ irrep 
and $\ell$ is the level of weight $(x_k)$ within this irrep. Matrix $G^0$ was 
introduced in the previous section. The transformation to the orthogonal
basis $\mid\!\!\tilde{0}_i\rangle$, (eq.(\ref{eq:Cdef})), then diagonalizes 
the zero weight subspace. Using eq.(\ref{eq:00}) we get
\be
\mid\!\! 000000\rangle %
                   = \frac{1}{\sqrt{78}}\: \sum_{i=1}^{78}    (-1)^{\ell+1} \,\mid\! w\rangle\!\mid\! \ol{w}\rangle ,
\ee
where the last sum runs over the complete weight system of the $(000001)$. 
After the phase redefinition of the even level states we would get each Clebsch-Gordan coefficient the same,
$ 1/\sqrt{78}$, as one would expect for the singlet in the product of two self-conjugate 78-dimensional irreps.

\section{Application to model building: operator ${\bf 27\otimes 78} \otimes \ol{\bf 27}$ }

As a simple application we have derived the explicit form of 
the singlet operator contained in {\bf 27}$\otimes${\bf 78}$\otimes\ol{\bf 27}$ in terms of
the Standard Model gauge group multiplets. We assume the standard
embedding of the Standard Model states into the {\bf 27} in $E_6$
as summarized in table \ref{t:27states}. States of the $\ol{\bf 27}$ and 
{\bf 78} are labeled in tables \ref{t:27barstates} and
\ref{t:78states}, respectively, according to the similarity of their
$SU(3)_c\otimes SU(2)_L$ structure with the {\bf 27} irrep.\footnote
{
It is expected that the states of the {\bf 78}
and $\ol{\bf 27}$, as well as
$T$, $T^c$, $N^c$, and $S$ of the {\bf 27}  acquire 
very heavy masses and that is why they have not been observed.
In $N>1$ supersymmetry, the {\bf 78} in this operator contains vector particles 
that may be identified with the observed gauge bosons. 
}
Labeling of the non-zero {\bf 78} weights 
includes subscripts which indicate an SO(10) irrep the state belongs to.

The tables include signs associated with each
Dynkin label. The signs result from the conventions used for 
the embedding of the subgroup chain
\be
E_6 \;\supset\; SO(10) \;\supset\; SU(5) \;\supset\; SU(3)_c\otimes SU(2)_L\otimes U(1)_Y.
\label{eq:sbgrp}
\ee
In particular, our conventions for the $SO(10)$ projections read
\bea
E_{-\xi_1}& =& - \left[ E_{-\alpha_2},[E_{-\alpha_3},E_{-\alpha_4}] \right],\nnmb \\ 
E_{-\xi_2}& =& E_{-\alpha_6},\nnmb \\ 
E_{-\xi_3}& =& E_{-\alpha_3}, \label{eq:proj_SO10}\\
E_{-\xi_4}& =& [E_{-\alpha_4},E_{-\alpha_5}],\nnmb \\ 
E_{-\xi_5}& =& [E_{-\alpha_2},E_{-\alpha_1}], \nnmb 
\eea
where the $E_{-\xi_i},\:(i=1,\ldots 5)$ are the $SO(10)$ ladder operators
and $\xi_i$'s are the simple roots of $SO(10)$.
Similarly, $SU(5)$ lowerings are projected out according to
\bea
E_{-\eta_1}& =& [E_{-\xi_2},E_{-\xi_1}], \nnmb \\ 
E_{-\eta_2}& =& [E_{-\xi_3},E_{-\xi_5}], \label{eq:proj_SU5}\\
E_{-\eta_3}& =& E_{-\xi_4},              \nnmb \\ 
E_{-\eta_4}& =& [E_{-\xi_3},E_{-\xi_2}], \nnmb 
\eea
and $SU(3)$ and $SU(2)$ projections satisfy
\bea 
E_{-\pi_1} &=& [E_{-\eta_2},E_{-\eta_1}], \nnmb \\ 
E_{-\pi_2} &=& [E_{-\eta_3},E_{-\eta_4}],  \label{eq:proj_SU3xSU2}\\
E_{-\rho}\:&=& [E_{-\eta_2},E_{-\eta_3}].   \nnmb 
\eea	   
We remark that these projections are consistent with the explicit form 
of the $C$ matrix in eq.(\ref{eq:Cij}) and with relations (13) in ref.\cite{abI}.
Clearly, the sign at the $SO(10)$ weights, $SU(5)$ weights, or
$SU(3)_c\otimes SU(2)_L$ weights in tables \ref{t:27states}--\ref{t:78states} 
is a relative sign with respect to the $E_6$ weights, and follows from our choice
of the subgroup embedding.
We remind the reader that we have started with simple lowering phase convention 
which was just overall (+) sign for any weight state obtained by lowering in $E_6$ 
(compare with the text below eq.(\ref{eq:dfnN})). We also assume that the same simple 
lowering phase convention applies to the construction of any weight system within the 
subgroups of $E_6$. Signs in tables \ref{t:27states}--\ref{t:78states} indicate
that the embedding induces a relative phase for the states of $E_6$ and the corresponding
states of its subgroups. On top of the {\em embedding} phase convention,
we now introduce a third set of phase conventions, which we call {\em physical}. These
combine with the former but are not taken into account in tables
\ref{t:27states}--\ref{t:78states}. In particular, our physical phase conventions
for states of the $SU(3)_c\otimes SU(2)_L$ irreps read:
\begin{description}
\item[(A)] 
           Each $SU(3)$ anti-triplet component with weight (1$\,$\1) has its phase
           redefined by multiplying the state by ($-1$).
\item[(B)] 
           Anti-doublets of the $SU(2)$ are formed as 
           $\left( \begin{array}{r} (\:\ol{1}) \\ -(\:1) \end{array}\right)$
           with an extra ($-$) sign at the lower component,
           as opposed to doublets which are simply labeled as 
           $\left( \begin{array}{c} (\:1) \\ (\:\ol{1}) \end{array}\right)$.
\item[(C)] 
           $SU(3)$ octet components with weights (2$\,$\1) and (1$\,$\2) 
           and the weight (2) component of an $SU(2)$ triplet 
           have their phases redefined by multiplying the corresponding states by ($-1$).
\item[(D)] 
           Assuming that
           the two (0$\,$0) weight states of the $SU(3)$ octet are projected to be 
           orthogonal to each other and one of them lies in the isospin direction,\footnote
{
An example of such a construction is the standard set of eight
Gell-Mann matrices $\lambda^a$, see {\em e.g.}, a review on group theory 
in ref.\cite{PDG_SU3}. Gluon field $\lambda^8 A^8$ corresponds to 
the (0$\,$0) weight state which is an isospin singlet. 
$\lambda^3 A^3$, a member of the isospin triplet, is the orthogonal 
(0$\,$0) weight state. This notation is used in eq.(\ref{eq:7827b27}).
}
           the isospin singlet state has its phase redefined by multiplying the
           state with ($-1$).
\item[(E)] 
           The phases of the $D^c$ states (in the ${\bf 27}$ and ${\bf 78}$) and $\ol{E^c}$ 
           states (in the ${\bf 78}$ and $\ol{\bf 27}$) are redefined by multiplying
           the corresponding states with ($-1$).
\end{description}

The phase conventions (A)-(D) make up for the simplicity of the lowering phase 
convention for the Standard Model subgroups. In fact, they could be
substituted by a more complicated lowering rules at the $SU(3)_c\otimes SU(2)_L$ level, or
at the $E_6$ level.
The advantage of our approach is that the make-up changes are only suggested
at the $SU(3)\otimes SU(2)$ level after the weight system of an $E_6$ irrep is obtained with
simple lowering phase convention, and thus the construction is more transparent. 
Note that rules (A) and (B) of our physical phase conventions are 
introduced to make the singlet in $\ol{\bf f}\,{\bf f}$ a symmetric
combination (trace) of states in fundamental irreps ${\bf f}$ and $\ol{\bf f}$ of
the $SU(3)$ or $SU(2)$. Similarly, rules (C) and (D) put the singlet
in $\ol{\bf f}\,{\bf A}\,{\bf f}$ into the form familiar to particle physics, with  
the interaction Lagrangian 
$\ol{\Psi}_f\,T^a A^{a}\,\Psi_f$, where $A$ is the gauge field transforming as an adjoint
irrep and $T^a$s are the Gell-Mann matrices of $SU(3)$ or Pauli matrices of $SU(2)$.
Finally, according to our rule (E), $D^c$ states change sign to make the down quark mass term 
of the same sign as the up quark, electron, and neutrino mass terms, and the 
phase of $\ol{E^c}$ is redefined to keep the singlet in the 
${\bf 27} \otimes \ol{\bf 27}$ with plus signs only (trace), in terms of the
particle states.

Next, we specify which two-dimensional multiplets of $SU(2)$ 
(see tables \ref{t:27states}--\ref{t:78states})
are going to be labeled as doublets and which as anti-doublets. 
In our notation, two-component states
           $H_d = \left( \begin{array}{c} H_d^- \\ H_d^0 \end{array}\right)$
           of the {\bf 27},  
$\ol{X}$ of the {\bf 78}, and  
$\ol{Q}   = \left( \begin{array}{c} \ol{U}     \\ \ol{D}     \end{array}\right)$,
$\ol{L}   = \left( \begin{array}{c} \ol{\nu}   \\ \ol{e^-}   \end{array}\right)$,
           and
$\ol{H_u} = \left( \begin{array}{c} \ol{H_u^+} \\ \ol{H_u^0} \end{array}\right)$
           of both the ${\bf 78}$ and $\ol{\bf 27}$ represent $SU(2)_L$ anti-doublets.
Any other two-dimensional multiplets of $SU(2)$ are assumed to be doublets.
Our $SU(2)$ contractions among doublets and anti-doublets 
are defined to be as simple as possible: two doublets and two
anti-doublets are contracted through the same matrix 
$\epsilon=\left( \begin{array}{rc} 0 & 1 \\
                                  -1 & 0 \end{array} \right) $,
while the contraction of a doublet and an anti-doublet does not depend on the ordering.
For instance,
$QL = Ue^- - D\nu$, $\ol{Q}H_d = \ol{U}H_d^0 - \ol{D}H_d^-$,
and $QH_d = H_d Q = UH_d^- + DH_d^0$.

With all the phase conventions included the explicit form of the singlet 
operator contained in ${\bf 27\otimes 78} \otimes \ol{\bf 27}$ takes the form
\bea
\lefteqn{  ({\bf 78} \otimes \ol{\bf 27} \otimes {\bf 27})_1  = }                                               
                                                            \label{eq:7827b27} \\   
&& \!\!\!\!\! =
    \frac{1}{\sqrt{3}} \, A^a
    \left\{  \ol{Q}\frac{\lambda^a}{2}Q - \ol{U^c} \frac{\lambda^{a\,*}}{2}U^c 
                                        - \ol{D^c} \frac{\lambda^{a\,*}}{2}D^c 
           + \ol{T}\frac{\lambda^a}{2}T - \ol{T^c} \frac{\lambda^{a\,*}}{2}T^c   \right\}  \nnmb \\
&&\!\!\!\! 
  + \frac{1}{\sqrt{3}} \, W^i 
    \left\{  \ol{Q}  \frac{\sigma^i}{2}Q   + \ol{L}  \frac{\sigma^ i    }{2}L 
           + \ol{H_u}\frac{\sigma^i}{2}H_u - \ol{H_d}\frac{\sigma^{i\,*}}{2}H_d  \right\}  \nnmb \\
&&\!\!\!\!
  + \frac{1}{\sqrt{180}}\, Y^0
    \left\{ \ol{Q}Q -4 \ol{U^c}U^c + 6 \ol{E^c}E^c + 2 \ol{D^c}D^c -3 \ol{L}L
       -2 \ol{T}T + 3 \ol{H_u}H_u +2\,\ol{T^c}T^c -3 \ol{H_d}H_d             \right\}      \nnmb \\
&&\!\!\!\! 
  + \frac{1}{\sqrt{120}} \chi^0
    \left\{ - \ol{Q}Q - \ol{U^c}U^c - \ol{E^c}E^c + 3\,( \ol{D^c}D^c + \ol{L}L) - 5\, \ol{N^c}N^c
       +2\,(\ol{T}T + \ol{H_u}H_u - \ol{T^c}T^c - \ol{H_d}H_d)               \right\}      \nnmb \\
&&\!\!\!\! 
  + \frac{1}{\sqrt{72}} \Psi^0 
    \left\{  \ol{Q}Q + \ol{U^c}U^c + \ol{E^c}E^c + \ol{D^c}D^c + \ol{L}L + \ol{N^c}N^c
       -2\,(\ol{T}T + \ol{H_u}H_u + \ol{T^c}T^c + \ol{H_d}H_d)  + 4\,\ol{S}S \right\}      \nnmb \\
&&\!\!\!\! 
  + \frac{1}{\sqrt{6}} X 
    \left\{ -\ol{Q}E^c + \ol{L}D^c - \ol{T}H_u + \ol{H_d}T^c - \ol{U^c}Q     \right\}      
\nnmb \\
 &&\!\!\!\! 
  + \frac{1}{\sqrt{6}} \ol{X} 
    \left\{ +\ol{E^c}Q - \ol{D^c}L + \ol{H_u}T - \ol{T^c}H_d - \ol{Q}U^c     \right\}      \nnmb \\
&&\!\!\!\! 
  + \frac{1}{\sqrt{6}} Q_{45}
    \left\{ -\ol{Q}N^c + \ol{L}U^c - \ol{H_u}T^c + \ol{T}H_d + \ol{D^c}Q     \right\}      \nnmb \\
&&\!\!\!\! 
  + \frac{1}{\sqrt{6}} U^c_{45}
    \left\{ -\ol{U^c}N^c - \ol{D^c}E^c + \ol{L}Q + \ol{T}T^c                 \right\}     
  + \frac{1}{\sqrt{6}} E^c_{45}
    \left\{ -\ol{E^c}N^c - \ol{H_u}H_d - \ol{D^c}U^c                         \right\}      \nnmb \\
&&\!\!\!\! 
  + \frac{1}{\sqrt{6}} \ol{Q}_{45} 
    \left\{ +\ol{N^c}Q - \ol{U^c}L + \ol{T^c}H_u - \ol{H_d}T + \ol{Q}D^c     \right\}      \nnmb \\
&&\!\!\!\! 
  + \frac{1}{\sqrt{6}} \ol{U^c}_{45}
    \left\{ +\ol{N^c}U^c + \ol{E^c}D^c - \ol{Q}L + \ol{T^c}T                 \right\}     
  + \frac{1}{\sqrt{6}} \ol{E^c}_{45}
    \left\{ +\ol{N^c}E^c - \ol{H_d}H_u + \ol{U^c}D^c                         \right\}      \nnmb \\
&&\!\!\!\! 
  + \frac{1}{\sqrt{6}} Q_{16}
    \left\{ \ol{Q}S + \ol{T}L + \ol{H_u}D^c + \ol{H_d}U^c + \ol{T^c}Q        \right\}     
  + \frac{1}{\sqrt{6}} U^c_{16}
    \left\{ \ol{U^c}S - \ol{T^c}E^c - \ol{H_d}Q - \ol{T}D^c                  \right\}      \nnmb \\
&&\!\!\!\! 
  + \frac{1}{\sqrt{6}} E^c_{16}
    \left\{ \ol{E^c}S + \ol{H_u}L - \ol{T^c}U^c                              \right\}     
  + \frac{1}{\sqrt{6}} D^c_{16}
    \left\{ \ol{D^c}S + \ol{T^c}N^c + \ol{H_u}Q + \ol{T}U^c                  \right\}      \nnmb \\
&&\!\!\!\! 
  + \frac{1}{\sqrt{6}} L_{16}
    \left\{ \ol{L}S + \ol{H_u}E^c + \ol{H_d}N^c - \ol{T}Q                    \right\}     
  + \frac{1}{\sqrt{6}} N^c_{16}
    \left\{ \ol{N^c}S - \ol{H_d}L + \ol{T^c}D^c                              \right\}      \nnmb \\
&&\!\!\!\! 
  + \frac{1}{\sqrt{6}} \ol{Q}_{\ol{16}} 
    \left\{ -\ol{S}Q - \ol{L}T - \ol{D^c}H_u - \ol{U^c}H_d + \ol{Q}T^c       \right\}     
  + \frac{1}{\sqrt{6}} \ol{U^c}_{\ol{16}} 
    \left\{ -\ol{S}U^c + \ol{E^c}T^c - \ol{Q}H_d - \ol{D^c}T                 \right\}      \nnmb \\
&&\!\!\!\! 
  + \frac{1}{\sqrt{6}} \ol{E^c}_{\ol{16}} 
    \left\{ -\ol{S}E^c - \ol{L}H_u + \ol{U^c}T^c                             \right\}     
  + \frac{1}{\sqrt{6}} \ol{D^c}_{\ol{16}} 
    \left\{ -\ol{S}D^c - \ol{N^c}T^c - \ol{Q}H_u + \ol{U^c}T                 \right\}      \nnmb \\
&&\!\!\!\! 
  + \frac{1}{\sqrt{6}} \ol{L}_{\ol{16}}  
    \left\{ -\ol{S}L - \ol{E^c}H_u - \ol{N^c}H_d + \ol{Q}T                   \right\}     
  + \frac{1}{\sqrt{6}} \ol{N^c}_{\ol{16}} 
    \left\{ -\ol{S}N^c - \ol{L}H_d - \ol{D^c}T^c                             \right\},     \nnmb 
\eea
where the orthogonal zero weight states have been obtained using matrix $C$ given in eq.(\ref{eq:Cij}).
$|\tilde{0}_4\rangle$, $|\tilde{0}_5\rangle$, and $|\tilde{0}_6\rangle$ are now labeled
as $Y^0$, $\chi^0$, and $\Psi^0$, respectively.
As usual in particle physics, ``gluon'' fields $A^a$ ($a=1,..,8$) are defined via relations 
$G_{(\pm1\pm1)}=(A^4\mp iA^5)/\sqrt{2}$, 
$G_{(\pm2\mp1)}=(A^1\mp iA^2)/\sqrt{2}$, 
$G_{(\mp1\pm2)}=(A^6\mp iA^7)/\sqrt{2}$, 
$ - |\tilde{0}_2\rangle\equiv G_{(00)}^{I-singlet} = A^8$, and
$   |\tilde{0}_1\rangle\equiv G_{(00)}^{I-triplet} = A^3$, and 
the $SU(2)$ triplet fields satisfy
$W_{(\pm2)} = (W^1\mp iW^2)/\sqrt{2}$, and
$|\tilde{0}_3\rangle \equiv W_{(0)} = W^3$.
Note that the $Y^0$, $\chi^0$, and $\Psi^0$ interaction terms
include numerical factors which coincide with 
the $Q^z$ (hypercharge, in the standard embedding which we follow in this paper),
$Q^r$, and $Q^t$ charges, respectively, of the components of the ${\bf 27}$ 
calculated in \cite{abI}. This provides an important check of our calculation. 
Another interesting detail is the antisymmetry 
between off-diagonal charge conjugated terms. This is a direct consequence 
of the conventions we use. A symmetric property could be restored by a
broader set of {\em physical} conventions. In fact, rule (C) of our physical phase conventions
does exactly that for the off-diagonal contractions containing the $SU(3)$ and $SU(2)$ adjoints.
Alternatively, we could start with a different {\em lowering} phase convention.

\section{Summary}

In this paper we calculated the Clebsch-Gordan decomposition of the
tensor product of two adjoints in $E_6$. In detail, we explained the steps 
related to the presence of degenerate zero weights
in the {\bf 78}. 
Our results can be applied to unification model building in a straightforward way. 
As a simple application we worked out a complete form
of the singlet ${\bf 27\otimes 78} \otimes \ol{\bf 27}$ operator. 
In addition, the decomposition
of the ${\bf 78} \otimes {\bf 78}$ tensor product may be useful
for a detailed study of the symmetry breaking sector of unified theories based on $E_6$, 
and for the analysis of higher dimensional operators in these theories, which contain fields 
transforming as an adjoint representation of $E_6$.

\protect
\begin{table}[t]
\caption{ 
{\bf Bases in the dominant weight subspaces 
of the 2430-dimensional $(000002)$ irrep.} 
\hspace*{8.0in}.  
(001000) weight, eq.(\ref{eq:001000s}), 
is left out as trivial.
\hspace*{8.0in}.  
\hspace*{8.0in}.
}
\label{t:paths_2430}
\begin{tabular}{|l|c||l|c|}
\hline
                                   &                          &                                   &                        \\
  Weight state                     &  Lowering path           &  Weight state                     &  Lowering path         \\
                                   &                          &                                   &                        \\
\hline 
                                   &                          &                                   &                        \\
$\;\;\mid\!\! 100010_1     \rangle$&  634236                  &$\;\;\mid\!\! 000000_{10}  \rangle$& 6364534523423412361236 \\
$\;\;\mid\!\! 100010_2     \rangle$&  364236                  &$\;\;\mid\!\! 000000_{11}  \rangle$& 6345236432112364534236 \\
$\;\;\mid\!\! 100010_3     \rangle$&  436236                  &$\;\;\mid\!\! 000000_{12}  \rangle$& 3366454523423412361236 \\
                                   &                          &$\;\;\mid\!\! 000000_{13}  \rangle$& 3516443222133456634236 \\
 \cline{1-2}                       &                          &$\;\;\mid\!\! 000000_{14}  \rangle$& 3562312444533216634236 \\
$\;\;\mid\!\! 000001_{\: 1}\rangle$& 63452341236              &$\;\;\mid\!\! 000000_{15}  \rangle$& 3562144533642213364236 \\
$\;\;\mid\!\! 000001_{\: 2}\rangle$& 36452341236              &$\;\;\mid\!\! 000000_{16}  \rangle$& 3164354222133456634236 \\
$\;\;\mid\!\! 000001_{\: 3}\rangle$& 35144236236              &$\;\;\mid\!\! 000000_{17}  \rangle$& 3164522133624453364236 \\
$\;\;\mid\!\! 000001_{\: 4}\rangle$& 34523641236              &$\;\;\mid\!\! 000000_{18}  \rangle$& 3644365523423412361236 \\
$\;\;\mid\!\! 000001_{\: 5}\rangle$& 52312436436              &$\;\;\mid\!\! 000000_{19}  \rangle$& 3643542234653412361236 \\
$\;\;\mid\!\! 000001_{\: 6}\rangle$& 54362341236              &$\;\;\mid\!\! 000000_{20}  \rangle$& 3645223364453412361236 \\
$\;\;\mid\!\! 000001_{\: 7}\rangle$& 14534236236              &$\;\;\mid\!\! 000000_{21}  \rangle$& 3645236432112364534236 \\
$\;\;\mid\!\! 000001_{\: 8}\rangle$& 12364534236              &$\;\;\mid\!\! 000000_{22}  \rangle$& 4436365523423412361236 \\
$\;\;\mid\!\! 000001_{\: 9}\rangle$& 42513634236              &$\;\;\mid\!\! 000000_{23}  \rangle$& 4251334521663434236236 \\
$\;\;\mid\!\! 000001_{10}  \rangle$& 43652341236              &$\;\;\mid\!\! 000000_{24}  \rangle$& 4251334236663452341236 \\
$\;\;\mid\!\! 000001_{11}  \rangle$& 23645341236              &$\;\;\mid\!\! 000000_{25}  \rangle$& 4153463222133456634236 \\
                                   &                          &$\;\;\mid\!\! 000000_{26}  \rangle$& 4153622133624453364236 \\
 \cline{1-2}                       &                          &$\;\;\mid\!\! 000000_{27}  \rangle$& 4365543623423412361236 \\
$\;\;\mid\!\! 000000_{\: 1}\rangle$& 6634534523423412361236   &$\;\;\mid\!\! 000000_{28}  \rangle$& 4365432234653412361236 \\
$\;\;\mid\!\! 000000_{\: 2}\rangle$& 6524133364452213364236   &$\;\;\mid\!\! 000000_{29}  \rangle$& 5543643623423412361236 \\
$\;\;\mid\!\! 000000_{\: 3}\rangle$& 6514233364452213364236   &$\;\;\mid\!\! 000000_{30}  \rangle$& 5241334521663434236236 \\
$\;\;\mid\!\! 000000_{\: 4}\rangle$& 6543364523423412361236   &$\;\;\mid\!\! 000000_{31}  \rangle$& 5241334236663452341236 \\
$\;\;\mid\!\! 000000_{\: 5}\rangle$& 6245133364452213364236   &$\;\;\mid\!\! 000000_{32}  \rangle$& 5143622133624453364236 \\
$\;\;\mid\!\! 000000_{\: 6}\rangle$& 6213362145453434236236   &$\;\;\mid\!\! 000000_{33}  \rangle$& 2236364545343412361236 \\
$\;\;\mid\!\! 000000_{\: 7}\rangle$& 6415233364452213364236   &$\;\;\mid\!\! 000000_{34}  \rangle$& 2361123645453434236236 \\
$\;\;\mid\!\! 000000_{\: 8}\rangle$& 6453364523423412361236   &$\;\;\mid\!\! 000000_{35}  \rangle$& 2361234432631254365436 \\
$\;\;\mid\!\! 000000_{\: 9}\rangle$& 6123362145453434236236   &$\;\;\mid\!\! 000000_{36}  \rangle$& 1123623645453434236236 \\
                                   &                          &                                   &                        \\
\hline					                    
\end{tabular}				                    
\end{table}				                    

\clearpage

\protect
\begin{table}[t]
\caption{ 
{\bf Bases in the dominant weight subspaces 
of the 2925-dimensional $(001000)$ irrep.} 
\hspace*{8.0in}.  
}
\label{t:paths_2925}
\begin{tabular}{|l|c||l|c|}
\hline
                                   &                          &                                   &                        \\
  Weight state                     &  Lowering path           &  Weight state                     &  Lowering path         \\
                                   &                          &                                   &                        \\
\hline 
                                   &                          &                                   &                        \\
$\;\;\mid\!\! 100010_1     \rangle$&  63423                   &$\;\;\mid\!\! 000000_{12}  \rangle$&  612321453632454363423 \\
$\;\;\mid\!\! 100010_2     \rangle$&  36423                   &$\;\;\mid\!\! 000000_{13}  \rangle$&  636453422345311236423 \\
$\;\;\mid\!\! 100010_3     \rangle$&  43623                   &$\;\;\mid\!\! 000000_{14}  \rangle$&  633442362554311236423 \\
$\;\;\mid\!\! 100010_4     \rangle$&  23643                   &$\;\;\mid\!\! 000000_{15}  \rangle$&  634523643212364534123 \\
                                   &                          &$\;\;\mid\!\! 000000_{16}  \rangle$&  524134532663241363423 \\
 \cline{1-2}                       &                          &$\;\;\mid\!\! 000000_{17}  \rangle$&  524134263453261363423 \\
$\;\;\mid\!\! 000001_{\: 1}\rangle$&  6345234123              &$\;\;\mid\!\! 000000_{18}  \rangle$&  524134321663452363423 \\
$\;\;\mid\!\! 000001_{\: 2}\rangle$&  3645234123              &$\;\;\mid\!\! 000000_{19}  \rangle$&  513644321223465363423 \\
$\;\;\mid\!\! 000001_{\: 3}\rangle$&  3145236423              &$\;\;\mid\!\! 000000_{20}  \rangle$&  514362134223465363423 \\
$\;\;\mid\!\! 000001_{\: 4}\rangle$&  3521436423              &$\;\;\mid\!\! 000000_{21}  \rangle$&  514362233643521436423 \\
$\;\;\mid\!\! 000001_{\: 5}\rangle$&  3452364123              &$\;\;\mid\!\! 000000_{22}  \rangle$&  536231245443261363423 \\
$\;\;\mid\!\! 000001_{\: 6}\rangle$&  5241363423              &$\;\;\mid\!\! 000000_{23}  \rangle$&  536214532443261363423 \\
$\;\;\mid\!\! 000001_{\: 7}\rangle$&  5142363423              &$\;\;\mid\!\! 000000_{24}  \rangle$&  536214436323145236423 \\
$\;\;\mid\!\! 000001_{\: 8}\rangle$&  5436234123              &$\;\;\mid\!\! 000000_{25}  \rangle$&  545342312663241363423 \\
$\;\;\mid\!\! 000001_{\: 9}\rangle$&  2451363423              &$\;\;\mid\!\! 000000_{26}  \rangle$&  544332266345311236423 \\
$\;\;\mid\!\! 000001_{10}  \rangle$&  2364534123              &$\;\;\mid\!\! 000000_{27}  \rangle$&  245134532663241363423 \\
$\;\;\mid\!\! 000001_{11}  \rangle$&  2345123643              &$\;\;\mid\!\! 000000_{28}  \rangle$&  245134263453261363423 \\
$\;\;\mid\!\! 000001_{12}  \rangle$&  4152363423              &$\;\;\mid\!\! 000000_{29}  \rangle$&  245134321663452363423 \\
$\;\;\mid\!\! 000001_{13}  \rangle$&  4365234123              &$\;\;\mid\!\! 000000_{30}  \rangle$&  211345234663452363423 \\
$\;\;\mid\!\! 000001_{14}  \rangle$&  4354236123              &$\;\;\mid\!\! 000000_{31}  \rangle$&  212334466332155436423 \\
$\;\;\mid\!\! 000001_{15}  \rangle$&  1236453423              &$\;\;\mid\!\! 000000_{32}  \rangle$&  213324466332155436423 \\
                                   &                          &$\;\;\mid\!\! 000000_{33}  \rangle$&  213214534663452363423 \\
 \cline{1-2}                       &                          &$\;\;\mid\!\! 000000_{34}  \rangle$&  213245346633452364123 \\
$\;\;\mid\!\! 000000_{\: 1}\rangle$&  652413633454221363423   &$\;\;\mid\!\! 000000_{35}  \rangle$&  415346321223465363423 \\
$\;\;\mid\!\! 000000_{\: 2}\rangle$&  651423633454221363423   &$\;\;\mid\!\! 000000_{36}  \rangle$&  415362134223465363423 \\
$\;\;\mid\!\! 000000_{\: 3}\rangle$&  654363422345311236423   &$\;\;\mid\!\! 000000_{37}  \rangle$&  415362233643521436423 \\
$\;\;\mid\!\! 000000_{\: 4}\rangle$&  654345213634221363423   &$\;\;\mid\!\! 000000_{38}  \rangle$&  453342266345311236423 \\
$\;\;\mid\!\! 000000_{\: 5}\rangle$&  624513633454221363423   &$\;\;\mid\!\! 000000_{39}  \rangle$&  453452312663241363423 \\
$\;\;\mid\!\! 000000_{\: 6}\rangle$&  621363244332155436423   &$\;\;\mid\!\! 000000_{40}  \rangle$&  453423126633452364123 \\
$\;\;\mid\!\! 000000_{\: 7}\rangle$&  621321453632454363423   &$\;\;\mid\!\! 000000_{41}  \rangle$&  136435421223465363423 \\
$\;\;\mid\!\! 000000_{\: 8}\rangle$&  641523633454221363423   &$\;\;\mid\!\! 000000_{42}  \rangle$&  136452134223465363423 \\
$\;\;\mid\!\! 000000_{\: 9}\rangle$&  645363422345311236423   &$\;\;\mid\!\! 000000_{43}  \rangle$&  136452233643521436423 \\
$\;\;\mid\!\! 000000_{10}  \rangle$&  645345213634221363423   &$\;\;\mid\!\! 000000_{44}  \rangle$&  364354234652311236423 \\
$\;\;\mid\!\! 000000_{11}  \rangle$&  612363244332155436423   &$\;\;\mid\!\! 000000_{45}  \rangle$&  364523643212364534123 \\
                                   &                          &                                   &                        \\
\hline					                    
\end{tabular}				                    
\end{table}				                    

\clearpage

\protect
\begin{table}[t]
\caption{ 
{\bf CG coefficients for (100010) dominant weight in (000001)$\otimes$(000001).}
\hspace*{8.0in}.  
Each entry should be divided by the respective number in the last row to keep
the degenerate 
\hspace*{8.0in}.  
states normalized to 1.
\hspace*{8.0in}.  
\hspace*{8.0in}.  
}
\label{t:cgc100010}
\footnotesize
\begin{tabular}{|@{\hspace{0.5mm}}c
                |c@{$\,$}c@{$\,$}c@{$\,$}
                |c@{$\,$}c@{$\,$}c@{$\,$}c@{$\,$}
                |c@{$\,$}|}
\hline
\mbox{ } & \multicolumn{3}{ c|}{\mbox{ }} & \multicolumn{4}{|c|}{\mbox{ }} & {\mbox{ }} \\
\mbox{ } & \multicolumn{3}{ c|}{\mbox{ }} & \multicolumn{4}{|c|}{\mbox{ }} & {\mbox{ }} \\
\mbox{ } & \multicolumn{3}{ c|}{\normalsize \em (000002) } & 
                                              \multicolumn{4}{|c|}{ \normalsize \em (001000) } &
                                                                   {\normalsize \em (100010) } \\
\mbox{ } & \multicolumn{3}{ c|}{\mbox{ }} & \multicolumn{4}{|c|}{\mbox{ }} & {\mbox{ }} \\
\mbox{ } & \multicolumn{3}{ c|}{\mbox{ }} & \multicolumn{4}{|c|}{\mbox{ }} & {\mbox{ }} \\
\cline{2-9}
                          &     &      &      &      &      &      &      &      \\
 & 
   $\,\mid$100010$_{1}\rangle$ & 
   $\,\mid$100010$_{2}\rangle$ & 
   $\,\mid$100010$_{3}\rangle$ & 
   $\,\mid$100010$_{1}\rangle$ & 
   $\,\mid$100010$_{2}\rangle$ & 
   $\,\mid$100010$_{3}\rangle$ & 
   $\,\mid$100010$_{4}\rangle$ & 
   $\,\mid$100010$        \rangle$ \\ 
                          &     &      &      &      &      &      &      &      \\
\hline
%                         &  1      2      3      1      2      3      4      1           
%                         ----------------------------------------------------------
                          &     &      &      &      &      &      &      &      \\
\AA10001\1\MM000001\ZZ    &  1  &      &      &   1  &      &      &      &   1  \\
                          &     &      &      &      &      &      &      &      \\
\AA000001\MM10001\1\ZZ    &  1  &      &      &  -1  &      &      &      &   1  \\
                          &     &      &      &      &      &      &      &      \\
\AA10\1011\MM00100\1\ZZ   &  1  &   1  &      &   1  &   1  &      &      &  -1  \\
                          &     &      &      &      &      &      &      &      \\
\AA00100\1\MM10\1011\ZZ   &  1  &   1  &      &  -1  &  -1  &      &      &  -1  \\
                          &     &      &      &      &      &      &      &      \\
\AA1\11\110\MM01\1100\ZZ  &     &   1  &   1  &      &   1  &   1  &   1  &   1  \\
                          &     &      &      &      &      &      &      &      \\
\AA01\1100\MM1\11\110\ZZ  &     &   1  &   1  &      &  -1  &  -1  &  -1  &   1  \\
                          &     &      &      &      &      &      &      &      \\
\AA1\10100\MM010\110\ZZ   &     &      &   1  &      &      &   1  &  -1  &  -1  \\
                          &     &      &      &      &      &      &      &      \\
\AA010\110\MM1\10100\ZZ   &     &      &   1  &      &      &  -1  &   1  &  -1  \\
                          &     &      &      &      &      &      &      &      \\
\hline
                          &     &      &      &      &      &      &      &      \\
                          &  2  &   2  &   2  &   2  &   2  &   2  &   2  & $\sqrt{8}$ \\
                          &     &      &      &      &      &      &      &      \\
\hline
\end{tabular}
\end{table}

\clearpage

\protect
\begin{table}
\caption{ 
{\bf CG coefficients for (000001) dominant weight in (000001)$\otimes$(000001).}
\hspace*{8.0in}.  
In the last column, $A$ stands for the adjoint $(000001)$ irrep. 
\AA n\ZZ\ is an abbreviation for \AA000001$_n$\ZZ. 
\hspace*{8.0in}.  
Each CGC should be divided by the respective number
in the last row to maintain $\langle n\!\mid\! n\rangle = 1$. 
\hspace*{8.0in}.  
}
\label{t:cgc000001}
\tiny
%                 12  34567890123  456789012345678  90123  4
%\begin{tabular}{|cc||rrrrrrrrrrr||rrrrrrrrrrrrrrr||rrrrr||r|}
\begin{tabular}{|@{\hspace{0.5mm}}l@{\hspace{0.5mm}}l@{\hspace{1mm}}
                |@{\hspace{0mm}}r@{}r@{}r@{}r@{}r@{}r@{}r@{}r@{}r@{}r@{}r@{}
                |@{\hspace{0mm}}r@{$\,$}r@{$\,$}r@{$\,$}r@{$\,$}r@{$\,$}r@{$\,$}r@{$\,$}r@{$\,$}r@{$\,$}r@{$\,$}
                                                            r@{$\,$}r@{$\,$}r@{$\,$}r@{$\,$}r@{$\,$}
                |@{\hspace{0mm}}r@{}r@{$\,$}r@{$\,$}r@{$\,$}r@{$\,$}
                |@{\hspace{0mm}}r@{\hspace{1.0mm}}|}
\hline
 & &  \multicolumn{11}{ c|}{\mbox{ }} & 
      \multicolumn{15}{|c|}{\mbox{ }} & 
      \multicolumn{ 5}{|c|}{\mbox{ }} & {\mbox{ }} \\
 & &  \multicolumn{11}{ c|}{\normalsize \em (000002) } & 
      \multicolumn{15}{|c|}{\normalsize \em (001000) } &
      \multicolumn{ 5}{|c|}{\normalsize \em (100010) } &
                           {\normalsize \em A} \\
 & &  \multicolumn{11}{ c|}{\mbox{ }} & 
      \multicolumn{15}{|c|}{\mbox{ }} & 
      \multicolumn{ 5}{|c|}{\mbox{ }} & {\mbox{ }} \\
\cline{3-34}
 & & 
    $\!\!$\BG 1\ZZ& \BG 2\ZZ& \BG 3\ZZ& \BG 4\ZZ& \BG 5\ZZ& \BG 6\ZZ& \BG 7\ZZ& \BG 8\ZZ& \BG 9\ZZ& \AA 10\ZZ& \AA 11\ZZ&
    $\!\!$\BG 1\ZZ& \BG 2\ZZ& \BG 3\ZZ& \BG 4\ZZ& \BG 5\ZZ& \BG 6\ZZ& \BG 7\ZZ& \BG 8\ZZ& \BG 9\ZZ& \AA 10\ZZ& \AA 11\ZZ& 
                                                                                                \AA 12\ZZ& \AA 13\ZZ& \AA 14\ZZ& \AA 15\ZZ&
$\!\!$\BG 1\ZZ& \BG 2\ZZ& \BG 3\ZZ& \BG 4\ZZ& \BG 5\ZZ& 
$\!\!$\BG 1\ZZ \\ 
\hline
%                         &  1   2   3   4   5   6   7   8   9   0   1*  1   2   3   4   5   6   7   8   9   0   1   2   3   4   5*   1   2   3   4   5*  1 
%                          ---------------------------------------------------------------------------------------------------------------------------------
            &             &   &   &   &   &   &   &   &   &   &   &   &    &   &   &   &   &   &   &   &   &   &   &   &   &   &   &   &   &   &   &   &   \\
\AA000001\ZZ&\NA          &   &   &   &   &   &  &&$\sqrt{2}$&&   &   &    &   &   &   &   &   &   &   &   &   &   &  &&&$\mqrt{2}$& $\sqrt{2}$&&&&& -$\sqrt{2}\;$\\
            &             &   &   &   &   &   &   &   &   &   &   &   &    &   &   &   &   &   &   &   &   &   &   &   &   &   &   &   &   &   &   &   &   \\
\mbmm\NA&\AA000001\ZZ     &   &   &   &   &   &  &&$\sqrt{2}$&&   &   &    &   &   &   &   &   &   &   &   &   &   &  &&&$\sqrt{2}$& $\sqrt{2}$&&&&&  $\sqrt{2}\;$\\
            &             &   &   &   &   &   &   &   &   &   &   &   &    &   &   &   &   &   &   &   &   &   &   &   &   &   &   &   &   &   &   &   &   \\
\AA000001\ZZ&\NB          &   &   &   &   &   &   &   &  &&&$\sqrt{2}$&    &   &   &   &   &   &   &  &&$\mqrt{2}$&&   &   &   &   & &$\sqrt{2}$&&&&  $\sqrt{8}\;$\\
            &             &   &   &   &   &   &   &   &   &   &   &   &    &   &   &   &   &   &   &   &   &   &   &   &   &   &   &   &   &   &   &   &   \\
\mbmm\NB&\AA000001\ZZ     &   &   &   &   &   &   &   &  &&&$\sqrt{2}$&    &   &   &   &   &   &   &  &&$\sqrt{2}$&&   &   &   &   & &$\sqrt{2}$&&&& -$\sqrt{8}\;$\\
            &             &   &   &   &   &   &   &   &   &   &   &   &    &   &   &   &   &   &   &   &   &   &   &   &   &   &   &   &   &   &   &   &   \\
\AA000001\ZZ&\NC          &&$\sqrt{2}$&&  &   &   &   &   &   &   &   &  &$\mqrt{2}$&& &   &   &   &   &   &   &   &   &   &   &   & &&$\sqrt{2}$&&& -$\sqrt{18}$\\
            &             &   &   &   &   &   &   &   &   &   &   &   &    &   &   &   &   &   &   &   &   &   &   &   &   &   &   &   &   &   &   &   &   \\
\mbmm\NC&\AA000001\ZZ     &&$\sqrt{2}$&&  &   &   &   &   &   &   &   &  &$\sqrt{2}$&& &   &   &   &   &   &   &   &   &   &   &   & &&$\sqrt{2}$&&&  $\sqrt{18}$\\
            &             &   &   &   &   &   &   &   &   &   &   &   &    &   &   &   &   &   &   &   &   &   &   &   &   &   &   &   &   &   &   &   &   \\
\AA000001\ZZ&\ND          &   &   &   &   &   &   &   &  &&$\sqrt{2}$&&    &   &   &   &   &   &   &   &   &   &  &&$\mqrt{2}$&&   & &&&$\sqrt{2}$&&  $\sqrt{8}\;$\\
            &             &   &   &   &   &   &   &   &   &   &   &   &    &   &   &   &   &   &   &   &   &   &   &   &   &   &   &   &   &   &   &   &   \\
\mbmm\ND&\AA000001\ZZ     &   &   &   &   &   &   &   &  &&$\sqrt{2}$&&    &   &   &   &   &   &   &   &   &   &  &&$\sqrt{2}$&&   & &&&$\sqrt{2}$&& -$\sqrt{8}\;$\\
            &             &   &   &   &   &   &   &   &   &   &   &   &    &   &   &   &   &   &   &   &   &   &   &   &   &   &   &   &   &   &   &   &   \\
\AA000001\ZZ&\NE          &   &   &   &  &&$\sqrt{2}$&&   &   &   &   &    &   &   &   &   &  &&$\mqrt{2}$&&   &   &   &   &   &   & &&&&$\sqrt{2}$& -$\sqrt{2}\;$\\
            &             &   &   &   &   &   &   &   &   &   &   &   &    &   &   &   &   &   &   &   &   &   &   &   &   &   &   &   &   &   &   &   &   \\
\mbmm\NE&\AA000001\ZZ     &   &   &   &  &&$\sqrt{2}$&&   &   &   &   &    &   &   &   &   &  &&$\sqrt{2}$&&   &   &   &   &   &   & &&&&$\sqrt{2}$&  $\sqrt{2}\;$\\
            &             &   &   &   &   &   &   &   &   &   &   &   &    &   &   &   &   &   &   &   &   &   &   &   &   &   &   &   &   &   &   &   &   \\
\AA000001\ZZ&\NF          &$\sqrt{2}$&&&  &   &   &   &   &   &   &   &-$\sqrt{2}$&&&  &   &   &   &   &   &   &   &   &   &   &   &    &   &   & &&  $\sqrt{8}\;$\\
            &             &   &   &   &   &   &   &   &   &   &   &   &    &   &   &   &   &   &   &   &   &   &   &   &   &   &   &   &   &   &   &   &   \\
\mbmm\NF&\AA000001\ZZ     &$\sqrt{2}$&&&  &   &   &   &   &   &   &   & $\sqrt{2}$&&&  &   &   &   &   &   &   &   &   &   &   &   &    &   &   & && -$\sqrt{8}\;$\\
            &             &   &   &   &   &   &   &   &   &   &   &   &    &   &   &   &   &   &   &   &   &   &   &   &   &   &   &   &   &   &   &   &   \\
\AA00100\1\ZA00\1002\ZZ   &\p1&\p1&   &\p1&   &   &   &   &   &   &   & \m1&\m1&   &   &\m1&   &   &   &   &   &   &   &   &   &   &    &   &\m1&   &   & \m1 \\ 
            &             &   &   &   &   &   &   &   &   &   &   &   &    &   &   &   &   &   &   &   &   &   &   &   &   &   &   &   &   &   &   &   &   \\
\AA00\1002\ZA00100\1\ZZ   &\p1&\p1&   &\p1&   &   &   &   &   &   &   & \p1&\p1&   &   &\p1&   &   &   &   &   &   &   &   &   &   &    &   &\m1&   &   & \p1 \\
            &             &   &   &   &   &   &   &   &   &   &   &   &    &   &   &   &   &   &   &   &   &   &   &   &   &   &   &   &   &   &   &   &   \\
\AA01\1100\ZA0\11\101\ZZ  &   &\p1&   &\p1&   &   &   &   &\p1&\p1&\p1&    &\m1&   &   &\m1&   &   &   &\m1&\m1&   &   &\m1&   &   &    &\m1&\m1&\m1&   & \p1 \\
            &             &   &   &   &   &   &   &   &   &   &   &   &    &   &   &   &   &   &   &   &   &   &   &   &   &   &   &   &   &   &   &   &   \\
\AA0\11\101\ZA01\1100\ZZ  &   &\p1&   &\p1&   &   &   &   &\p1&\p1&\p1&    &\p1&   &   &\p1&   &   &   &\p1&\p1&   &   &\p1&   &   &    &\m1&\m1&\m1&   & \m1 \\ 
            &             &   &   &   &   &   &   &   &   &   &   &   &    &   &   &   &   &   &   &   &   &   &   &   &   &   &   &   &   &   &   &   &   \\
\AA1\10100\ZA\110\101\ZZ  &   &   &   &   &   &   &\p1&\p1&\p1&   &\p1&    &   &   &   &   &   &   &   &\m1&\m1&   &\m1&   &\m1&\m1&\mwo&\m1&   &\p1&   & \m1 \\ 
            &             &   &   &   &   &   &   &   &   &   &   &   &    &   &   &   &   &   &   &   &   &   &   &   &   &   &   &   &   &   &   &   &   \\
\AA\110\101\ZA1\10100\ZZ  &   &   &   &   &   &   &\p1&\p1&\p1&   &\p1&    &   &   &   &   &   &   &   &\p1&\p1&   &\p1&   &\p1&\p1&\mwo&\m1&   &\p1&   & \p1 \\
            &             &   &   &   &   &   &   &   &   &   &   &   &    &   &   &   &   &   &   &   &   &   &   &   &   &   &   &   &   &   &   &   &   \\
\AA010\110\ZA0\101\11\ZZ  &   &   &   &   &\p1&\p1&   &   &\p1&\p1&   &    &   &   &   &   &\m1&   &\m1&\m1&   &\m1&   &\m1&   &   &    &\p1&   &\m1&\m1& \m1 \\ 
            &             &   &   &   &   &   &   &   &   &   &   &   &    &   &   &   &   &   &   &   &   &   &   &   &   &   &   &   &   &   &   &   &   \\
\AA0\101\11\ZA010\110\ZZ  &   &   &   &   &\p1&\p1&   &   &\p1&\p1&   &    &   &   &   &   &\p1&   &\p1&\p1&   &\p1&   &\p1&   &   &    &\p1&   &\m1&\m1& \p1 \\
            &             &   &   &   &   &   &   &   &   &   &   &   &    &   &   &   &   &   &   &   &   &   &   &   &   &   &   &   &   &   &   &   &   \\
\AA\100100\ZA100\101\ZZ   &   &   &   &   &   &   &\p1&\p1&   &   &   &    &   &   &   &   &   &   &   &   &   &   &\m1&   &\p1&\m1&\mwo&   &   &\m1&   & \p1 \\ 
            &             &   &   &   &   &   &   &   &   &   &   &   &    &   &   &   &   &   &   &   &   &   &   &   &   &   &   &   &   &   &   &   &   \\
\AA100\101\ZA\100100\ZZ   &   &   &   &   &   &   &\p1&\p1&   &   &   &    &   &   &   &   &   &   &   &   &   &   &\p1&   &\m1&\p1&\mwo&   &   &\m1&   & \m1 \\
            &             &   &   &   &   &   &   &   &   &   &   &   &    &   &   &   &   &   &   &   &   &   &   &   &   &   &   &   &   &   &   &   &   \\
\AA1\11\110\ZA\11\11\11\ZZ&   &   &\p1&\p1&\p1&   &\p1&   &\p1&   &   &    &   &\m1&\m1&\m1&\m1&\m1&   &\m1&   &\m1&\m1&   &\m1&   & \p1&\p1&\p1&\p1&\p1& \p1 \\ 
            &             &   &   &   &   &   &   &   &   &   &   &   &    &   &   &   &   &   &   &   &   &   &   &   &   &   &   &   &   &   &   &   &   \\
\AA\11\11\11\ZA1\11\110\ZZ&   &   &\p1&\p1&\p1&   &\p1&   &\p1&   &   &    &   &\p1&\p1&\p1&\p1&\p1&   &\p1&   &\p1&\p1&   &\p1&   & \p1&\p1&\p1&\p1&\p1& \m1 \\
            &             &   &   &   &   &   &   &   &   &   &   &   &    &   &   &   &   &   &   &   &   &   &   &   &   &   &   &   &   &   &   &   &   \\
\AA0100\10\ZA0\10011\ZZ   &   &   &   &   &\p1&\p1&   &   &   &   &   &    &   &   &   &   &\m1&   &\m1&   &   &\p1&   &   &   &   &    &\m1&   &   &\m1& \p1 \\ 
            &             &   &   &   &   &   &   &   &   &   &   &   &    &   &   &   &   &   &   &   &   &   &   &   &   &   &   &   &   &   &   &   &   \\
\AA0\10011\ZA0100\10\ZZ   &   &   &   &   &\p1&\p1&   &   &   &   &   &    &   &   &   &   &\p1&   &\p1&   &   &\m1&   &   &   &   &    &\m1&   &   &\m1& \m1 \\
            &             &   &   &   &   &   &   &   &   &   &   &   &    &   &   &   &   &   &   &   &   &   &   &   &   &   &   &   &   &   &   &   &   \\
\AA\101\110\ZA10\11\11\ZZ &   &   &\p1&   &   &   &\p1&   &   &   &   &    &   &\m1&\p1&   &   &\m1&   &   &   &   &\m1&   &\p1&   & \p1&   &\m1&\m1&\m1& \m1 \\
            &             &   &   &   &   &   &   &   &   &   &   &   &    &   &   &   &   &   &   &   &   &   &   &   &   &   &   &   &   &   &   &   &   \\
\AA10\11\11\ZA\101\110\ZZ &   &   &\p1&   &   &   &\p1&   &   &   &   &    &   &\p1&\m1&   &   &\p1&   &   &   &   &\p1&   &\m1&   & \p1&   &\m1&\m1&\m1& \p1 \\
            &             &   &   &   &   &   &   &   &   &   &   &   &    &   &   &   &   &   &   &   &   &   &   &   &   &   &   &   &   &   &   &   &   \\
\AA10\1011\ZA\1010\10\ZZ  &   &   &\p1&\p1&   &   &   &   &   &   &   &    &   &\m1&\m1&\m1&   &\p1&   &   &   &   &   &   &   &   &\mwo&   &\p1&   &\m1& \m1 \\
            &             &   &   &   &   &   &   &   &   &   &   &   &    &   &   &   &   &   &   &   &   &   &   &   &   &   &   &   &   &   &   &   &   \\
\AA\1010\10\ZA10\1011\ZZ  &   &   &\p1&\p1&   &   &   &   &   &   &   &    &   &\p1&\p1&\p1&   &\m1&   &   &   &   &   &   &   &   &\mwo&   &\p1&   &\m1& \p1 \\
            &             &   &   &   &   &   &   &   &   &   &   &   &    &   &   &   &   &   &   &   &   &   &   &   &   &   &   &   &   &   &   &   &   \\
\AA1\110\10\ZA\11\1011\ZZ &   &   &\p1&   &\p1&   &   &   &   &   &   &    &   &\p1&\m1&   &\m1&\m1&   &   &   &\p1&   &   &   &   &\mwo&\m1&\m1&   &\p1& \m1 \\
            &             &   &   &   &   &   &   &   &   &   &   &   &    &   &   &   &   &   &   &   &   &   &   &   &   &   &   &   &   &   &   &   &   \\
\AA\11\1011\ZA1\110\10\ZZ &   &   &\p1&   &\p1&   &   &   &   &   &   &    &   &\m1&\p1&   &\p1&\p1&   &   &   &\m1&   &   &   &   &\mwo&\m1&\m1&   &\p1& \p1 \\
            &             &   &   &   &   &   &   &   &   &   &   &   &    &   &   &   &   &   &   &   &   &   &   &   &   &   &   &   &   &   &   &   &   \\
\hline			    
            &             &   &   &   &   &   &   &   &   &   &   &   &    &   &   &   &   &   &   &   &   &   &   &   &   &   &   &   &   &   &   &   &   \\
       &        & $\sqrt{6}$& $\sqrt{8}$& $\sqrt{8}$& $\sqrt{8}$& $\sqrt{8}$& $\sqrt{8}$& $\sqrt{8}$& $\sqrt{8}$& $\sqrt{8}$& $\sqrt{8}$& $\sqrt{8}$& 
                  $\sqrt{6}$& $\sqrt{8}$& $\sqrt{8}$& $\sqrt{8}$& $\sqrt{8}$& $\sqrt{8}$& $\sqrt{8}$& $\sqrt{8}$& $\sqrt{8}$& $\sqrt{8}$&
                                                                              $\sqrt{8}$& $\sqrt{8}$& $\sqrt{8}$& $\sqrt{8}$& $\sqrt{8}$&
                              4$\;$ & 4$\;$ & 4$\;$ & 4$\;$ & 4$\;$ &
                              $\sqrt{24}\;$  \\
            &             &   &   &   &   &   &   &   &   &   &   &   &    &   &   &   &   &   &   &   &   &   &   &   &   &   &   &   &   &   &   &   &   \\
\hline
\end{tabular}
\end{table}

\clearpage

\protect
\begin{table}[t]
\caption{ 
{\bf CG coefficients for (000000) dominant weight states of the 2430-dimensional 
\hspace*{8.0in}.  
(000002) irrep in the product (000001)$\otimes$(000001).}
\AA n\ZZ\ is an abbreviation for \AA000000$_n$\ZZ. 
\hspace*{8.0in}.  
Each CGC should be divided by the respective number in the last row to maintain
$\langle n\!\mid\! n\rangle = 1$.
\hspace*{8.0in}.  
}
\label{t:cgc000000_2430}
\tiny
%                            *         *         *         
%                 12  345678901234567890123456789012345678
%\begin{tabular}{|cc||rrrrrrrrrrrrrrrrrrrrrrrrrrrrrrrrrrrr|}
\begin{tabular}{|@{\hspace{0.5mm}}l@{\hspace{0.5mm}}l@{\hspace{1mm}}
                |@{\hspace{0mm}}c@{}c@{}c@{}c@{}c@{}c@{}c@{}c@{}c@{}c@{}
                                c@{}c@{}c@{}c@{}c@{}c@{}c@{}c@{}c@{}c@{}
                                c@{}c@{}c@{}c@{}c@{}c@{}c@{}c@{}c@{}c@{}
                                c@{}c@{}c@{}c@{}c@{}
                                                            c@{}|}
%                                                            c@{\hspace{1.0mm}}|}
\hline
 & &  \multicolumn{36}{ c|}{\mbox{                  }} \\ 
 & &  \multicolumn{36}{ c|}{\normalsize \em (000002) } \\
 & &  \multicolumn{36}{ c|}{\mbox{                  }} \\ 
\cline{3-38}
 & & 
          \AA 1\ZZ& \AA 2\ZZ& \AA 3\ZZ& \AA 4\ZZ& \AA 5\ZZ& \AA 6\ZZ& \AA 7\ZZ& \AA 8\ZZ& \AA 9\ZZ& \AA 10\ZZ& 
 \AA 11\ZZ& \AA 12\ZZ& \AA 13\ZZ& \AA 14\ZZ& \AA 15\ZZ& \AA 16\ZZ& \AA 17\ZZ& \AA 18\ZZ& \AA 19\ZZ& \AA 20\ZZ&
 \AA 21\ZZ& \AA 22\ZZ& \AA 23\ZZ& \AA 24\ZZ& \AA 25\ZZ& \AA 26\ZZ& \AA 27\ZZ& \AA 28\ZZ& \AA 29\ZZ& \AA 30\ZZ&
 \AA 31\ZZ& \AA 32\ZZ& \AA 33\ZZ& \AA 34\ZZ& \AA 35\ZZ& \AA 36\ZZ \\
\hline
%                         &1  2  3  4  5  6  7  8  9  0* 1  2  3  4  5  6  7  8  9  0* 1  2  3  4  5  6  7  8  9  0* 1  2  3  4  5  6
%                          -------------------------------------------------------------------------------------------------------------
        &                 &  &  &  &  &  &  &  &  &  &  &  &  &  &  &  &  &  &  &  &  &  &  &  &  &  &  &  &  &  &  &  &  &  &  &  &  \\
\mbmm\NA&\NA              &  &  &  &  &  &  &  &  &  &  &  &  &  &  &  &  &  &  &  &  &  &  &  &  &  &  &  &  &  &  &  &  &  &  &  &2 \\
\mbmm\NB&\NB              &  &  &  &  &  &  &  &  &  &  &  &  &  &  &  &  &  &  &  &  &  &  &  &  &  &  &  &  &  &  &  &  & 2&  & 4&  \\
\mbmm\NC&\NC              &  &  &  &  &  &  &  &  &  &  &  & 2&  &  &  &  &  &  & 4&  &  &  &  &  &  &  &  &  &  &  &  &  &  &  &  &  \\
\mbmm\ND&\ND              &  &  &  &  &  &  &  &  &  &  &  &  &  &  &  &  &  &  &  &  &  & 2&  &  &  &  &  & 4&  &  &  &  &  &  &  &  \\
\mbmm\NE&\NE              &  &  &  &  &  &  &  &  &  &  &  &  &  &  &  &  &  &  &  &  &  &  &  &  &  &  &  &  & 2&  &  &  &  &  &  &  \\
\mbmm\NF&\NF              & 2&  &  &  &  &  &  &  &  & 4& 4&  &  &  &  &  &  &  &  &  &  &  &  &  &  &  &  &  &  &  &  &  &  &  &  &  \\
        &                 &  &  &  &  &  &  &  &  &  &  &  &  &  &  &  &  &  &  &  &  &  &  &  &  &  &  &  &  &  &  &  &  &  &  &  &  \\
\mbmm\NA&\NB              &  &  &  &  &  &  &  &  &  &  &  &  &  &  &  &  &  &  &  &  &  &  &  &  &  &  &  &  &  &  &  &  &  & 2&  &  \\
\mbmm\NA&\NC              &  &  &  &  &  &  &  &  &  &  &  &  &  &  &  &  & 2&  &  &  &  &  &  &  &  &  &  &  &  &  &  &  &  &  &  &  \\
\mbmm\NA&\ND              &  &  &  &  &  &  &  &  &  &  &  &  &  &  &  &  &  &  &  &  &  &  &  &  &  & 2&  &  &  &  &  &  &  &  &  &  \\
\mbmm\NA&\NE              &  &  &  &  &  &  &  &  &  &  &  &  &  &  &  &  &  &  &  &  &  &  &  &  &  &  &  &  &  &  &  & 2&  &  &  &  \\
\mbmm\NA&\NF              &  &  &  &  &  &  &  &  & 2&  &  &  &  &  &  &  &  &  &  &  &  &  &  &  &  &  &  &  &  &  &  &  &  &  &  &  \\
\mbmm\NB&\NC              &  &  &  &  &  &  &  &  &  &  &  &  &  &  &  &  &  &  &  & 2&  &  &  &  &  &  &  &  &  &  &  &  &  &  &  &  \\
\mbmm\NB&\ND              &  &  &  &  &  &  &  &  &  &  &  &  &  &  &  &  &  &  &  &  &  &  & 2&  &  &  &  &  &  &  &  &  &  &  &  &  \\
\mbmm\NB&\NE              &  &  &  &  &  &  &  &  &  &  &  &  &  &  &  &  &  &  &  &  &  &  &  &  &  &  &  &  &  & 2&  &  &  &  &  &  \\
\mbmm\NB&\NF              &  &  &  &  &  & 2&  &  &  &  &  &  &  &  &  &  &  &  &  &  &  &  &  &  &  &  &  &  &  &  &  &  &  &  &  &  \\
\mbmm\NC&\ND              &  &  &  &  &  &  &  &  &  &  &  &  &  &  &  &  &  & 2&  &  &  &  &  &  &  &  &  &  &  &  &  &  &  &  &  &  \\
\mbmm\NC&\NE              &  &  &  &  &  &  &  &  &  &  &  &  &  &  & 2&  &  &  &  &  &  &  &  &  &  &  &  &  &  &  &  &  &  &  &  &  \\
\mbmm\NC&\NF              &  &  &  &  &  &  &  &  &  & 2&  &  &  &  &  &  &  &  &  &  & 2&  &  &  &  &  &  &  &  &  &  &  &  &  &  &  \\
\mbmm\ND&\NE              &  &  &  &  &  &  &  &  &  &  &  &  &  &  &  &  &  &  &  &  &  &  &  &  &  &  & 2&  &  &  &  &  &  &  &  &  \\
\mbmm\ND&\NF              &  &  &  &  &  &  &  & 2&  &  &  &  &  &  &  &  &  &  &  &  &  &  &  &  &  &  &  &  &  &  &  &  &  &  &  &  \\
\mbmm\NE&\NF              &  &  &  & 2&  &  &  &  &  &  &  &  &  &  &  &  &  &  &  &  &  &  &  &  &  &  &  &  &  &  &  &  &  &  &  &  \\
\AA000001\ZA00000\1\ZZ    &  &  &  &  &  &  &  &  &  &  & 1&  &  &  &  &  &  &  &  &  &  &  &  &  &  &  &  &  &  &  &  &  &  &  &  &  \\
\AA00100\1\ZA00\1001\ZZ   &  &  &  &  &  &  &  &  &  &  & 1&  &  &  &  &  &  &  &  &  & 1&  &  &  &  &  &  &  &  &  &  &  &  &  &  &  \\
\AA01\1100\ZA0\11\100\ZZ  &  &  &  &  &  &  &  &  &  &  &  &  &  &  &  &  &  &  &  &  & 1&  &  & 1&  &  &  &  &  &  &  &  &  &  &  &  \\
\AA1\10100\ZA\110\100\ZZ  &  &  &  &  &  &  &  &  &  &  &  &  &  &  &  &  &  &  &  &  &  &  &  & 1& 1&  &  &  &  &  &  &  &  &  &  &  \\
\AA010\110\ZA0\101\10\ZZ  &  &  &  &  &  &  &  &  &  &  &  &  &  &  &  &  &  &  &  &  &  &  &  & 1&  &  &  &  &  &  & 1&  &  &  &  &  \\ 
\AA\100100\ZA100\101\ZZ   &  &  &  &  &  &  &  &  &  &  &  &  &  &  &  &  &  &  &  &  &  &  &  &  & 1&  &  & 1&  &  &  &  &  &  &  &  \\ 
\AA1\11\110\ZA\11\11\11\ZZ&  &  &  &  &  &  &  &  &  &  &  &  & 1&  &  &  &  &  &  &  & 1&  &  & 1& 1&  &  &  &  &  & 1&  &  &  &  &  \\ 
\AA0100\10\ZA0\10011\ZZ   &  &  &  &  &  &  &  &  &  &  &  &  &  &  &  &  &  &  &  &  &  &  &  &  &  &  &  &  &  &  & 1&  &  &  & 1&  \\ 
%5:
\AA\101\110\ZA10\11\11\ZZ &  &  &  &  &  &  &  &  &  &  &  &  & 1& 1&  &  &  &  &  &  &  &  &  &  & 1&  &  & 1&  &  &  &  &  &  &  &  \\
\AA10\1011\ZA\1010\10\ZZ  &  &  & 1&  &  &  &  &  &  &  & 1&  & 1&  &  &  &  &  &  &  & 1&  &  &  &  &  &  &  &  &  &  &  &  &  &  &  \\
\AA1\110\10\ZA\11\1011\ZZ &  &  &  &  &  &  &  &  &  &  &  &  & 1&  &  & 1&  &  &  &  &  &  &  &  &  &  &  &  &  &  & 1&  &  &  & 1&  \\
%6:
\AA\1010\10\ZA10\1010\ZZ  &  &  &  &  &  &  &  &  &  &  &  &  & 1& 1&  & 1&  &  & 1&  & 1&  &  &  &  &  &  &  &  &  &  &  &  &  &  &  \\
\AA\11\1011\ZA1\110\1\1\ZZ&  & 1& 1&  &  &  &  &  &  &  &  &  & 1& 1&  &  &  &  &  &  &  &  &  &  &  &  &  &  &  &  & 1&  &  &  &  &  \\
\AA10001\1\ZA\1000\11\ZZ  &  &  & 1&  &  &  &  &  &  &  & 1&  &  &  &  &  &  &  &  &  &  &  &  &  &  &  &  &  &  &  &  & 1&  &  &  &  \\ 
\AA10\11\11\ZA\101\11\1\ZZ&  &  & 1&  &  &  & 1&  &  &  &  &  & 1&  &  & 1&  &  &  &  &  &  &  &  & 1&  &  &  &  &  &  &  &  &  &  &  \\ 
%7:
\AA\11\11\11\ZA1\11\11\1\ZZ& & 1& 1&  & 1&  & 1&  &  &  & 1&  & 1& 1&  & 1&  &  & 1&  & 1&  &  & 1& 1&  &  & 1&  &  & 1&  &  &  & 1&  \\
\AA0\10011\ZA0100\1\1\ZZ  &  & 1&  & 1&  &  &  &  &  &  &  &  &  &  &  &  &  &  &  &  &  &  &  &  &  &  &  &  &  &  & 1&  &  &  &  &  \\
\AA\11001\1\ZA1\100\11\ZZ &  & 1& 1&  &  &  &  &  &  &  &  &  &  &  &  &  &  &  &  &  &  &  &  &  &  &  &  &  &  & 1&  & 1&  &  &  &  \\
\AA1001\1\1\ZA\100\111\ZZ &  &  & 1&  &  &  & 1&  &  &  &  &  &  &  &  &  &  &  &  &  &  &  &  &  &  & 1&  &  &  &  &  & 1&  &  &  &  \\
\AA100\101\ZA\10010\1\ZZ  &  &  &  &  &  &  & 1&  & 1&  &  &  &  &  &  &  &  &  &  &  &  &  &  &  & 1&  &  &  &  &  &  &  &  &  &  &  \\
%8:
\AA0\101\11\ZA010\11\1\ZZ &  & 1&  & 1& 1&  &  & 1&  &  &  &  &  &  &  &  &  &  &  &  &  &  &  & 1&  &  &  &  &  &  & 1&  &  &  & 1&  \\
\AA\110\101\ZA1\1010\1\ZZ &  &  &  &  & 1& 1& 1&  & 1&  &  &  &  &  &  &  &  &  &  &  &  &  &  & 1& 1&  &  & 1&  &  &  &  &  &  &  &  \\
\AA\1101\1\1\ZA1\10\111\ZZ&  & 1& 1&  & 1&  & 1&  &  &  & 1&  &  &  &  &  &  &  &  &  &  &  & 1&  &  & 1&  &  &  & 1&  & 1&  &  &  &  \\
\AA0\1101\1\ZA01\10\11\ZZ &  & 1&  & 1&  &  &  &  &  &  &  &  &  & 1& 1&  &  &  &  &  &  &  &  &  &  &  &  &  &  & 1&  &  &  &  &  &  \\
\AA101\10\1\ZA\10\1101\ZZ &  &  &  &  &  &  & 1&  & 1&  &  &  &  &  &  & 1& 1&  &  &  &  &  &  &  &  & 1&  &  &  &  &  &  &  &  &  &  \\
%9:
\AA0\11\101\ZA01\110\1\ZZ &  &  &  &  & 1& 1&  & 1&  & 1& 1&  &  &  &  &  &  &  & 1&  & 2&  &  & 1&  &  &  &  &  &  &  &  &  &  &  &  \\
\AA0\111\1\1\ZA01\1\111\ZZ&  & 1&  & 1& 1&  &  & 1&  &  &  &  &  & 1& 1&  &  & 1& 1&  &  &  & 1&  &  &  &  & 1&  & 1&  &  &  &  &  &  \\
\AA\111\10\1\ZA1\1\1101\ZZ&  &  &  &  & 1& 1& 1&  & 1&  &  &  &  &  &  & 1& 1&  & 1& 1&  &  & 1&  &  & 1&  &  &  &  &  &  &  &  & 1&  \\
\AA00\1110\ZA001\1\10\ZZ  &  &  &  &  &  &  &  &  &  &  &  &  &  & 1& 1&  &  &  &  &  &  &  &  &  &  &  & 1& 1&  &  &  &  &  &  &  &  \\
\AA11\1000\ZA\1\11000\ZZ  &  &  &  &  &  &  &  &  &  &  &  &  &  &  &  & 1& 1&  &  &  &  &  &  &  &  &  &  &  &  &  &  &  &  & 1& 1&  \\
%10:
\AA2\10000\ZA\210000\ZZ   &  &  &  &  &  &  &  &  &  &  &  &  &  &  &  &  &  &  &  &  &  &  &  &  &  &  &  &  &  &  &  &  &  & 1& 1&1 \\
\AA\12\1000\ZA1\21000\ZZ  &  &  &  &  &  &  &  &  &  &  &  &  &  &  &  & 1& 1&  & 1& 1&  &  &  &  &  &  &  &  &  &  &  &  & 1& 1&  &  \\
\AA0\12\10\1\ZA01\2101\ZZ &  &  &  &  & 1& 1&  & 1&  & 1& 1& 1&  &  &  &  &  & 1& 2& 1& 1&  & 1&  &  &  &  & 1&  &  &  &  &  &  & 1&  \\
\AA00\12\10\ZA001\210\ZZ  &  &  &  &  &  &  &  &  &  &  &  &  &  & 1& 1&  &  & 1& 1&  &  & 1&  &  &  &  & 1& 2&  &  &  &  &  &  &  &  \\
\AA000\120\ZA0001\20\ZZ   &  &  &  &  &  &  &  &  &  &  &  &  &  &  &  &  &  &  &  &  &  &  &  &  &  &  & 1& 1& 1&  &  &  &  &  &  &  \\
\AA00\1002\ZA00100\2\ZZ   & 1&  &  &  &  &  &  &  &  & 3& 2&  &  &  &  &  &  &  & 1&  & 2&  &  &  &  &  &  &  &  &  &  &  &  &  &  &  \\
       &                  &  &  &  &  &  &  &  &  &  &  &  &  &  &  &  &  &  &  &  &  &  &  &  &  &  &  &  &  &  &  &  &  &  &  &  &  \\
%                          -------------------------------------------------------------------------------------------------------------
%                          &1  2  3  4  5  6  7  8  9  0* 1  2  3  4  5  6  7  8  9  0* 1  2  3  4  5  6  7  8  9  0* 1  2  3  4  5  6
\hline			    
       &                  &  &  &  &  &  &  &  &  &  &  &  &  &  &  &  &  &  &  &  &  &  &  &  &  &  &  &  &  &  &  &  &  &  &  &  &  \\
       &        & $\sqrt{6}$ &        4  &        4  &    4  &    4  &    4  &    4  &          4  &          4  &    8  &
                  $\sqrt{40}$& $\sqrt{6}$&        4  &    4  &    4  &    4  &    4  &          4  & $\sqrt{40}$ &    4  &
                  $\sqrt{40}$& $\sqrt{6}$&        4  &    4  &    4  &    4  &    4  & $\sqrt{40}$ & $\sqrt{ 6}$ &    4  &
                          4  &        4  & $\sqrt{6}$&    4  & 
                  $\sqrt{40}$& $\sqrt{ 6}\;$  \\
       &                  &  &  &  &  &  &  &  &  &  &  &  &  &  &  &  &  &  &  &  &  &  &  &  &  &  &  &  &  &  &  &  &  &  &  &  &  \\
\hline
\end{tabular}
\end{table}

\clearpage

\protect
\begin{table}[t]
\caption{ 
{\bf CG coefficients for the first 36 (000000) dominant weight states of 
     the 
\hspace*{8.0in}.  
2925-dimensional (001000) irrep in the product (000001)$\otimes$(000001).}
(The remaining 9 
\hspace*{8.0in}.  
states of this irrep with the same weight are shown in table \ref{t:cgc000000_2925_650_78_S}.)
\AA n\ZZ\   stands for \AA000000$_n$\ZZ. 
\hspace*{8.0in}.  
Each CGC should be divided by the respective number in the last row to maintain
$\langle n\!\mid\! n\rangle = 1$.
\hspace*{8.0in}.  
}
\label{t:cgc000000_2925}
\tiny
%                            *         *         *         
%                 12  345678901234567890123456789012345678
%\begin{tabular}{|cc||rrrrrrrrrrrrrrrrrrrrrrrrrrrrrrrrrrrr|}
\begin{tabular}{|@{\hspace{0.5mm}}l@{\hspace{0.5mm}}l@{\hspace{1mm}}
                |@{\hspace{0mm}}c@{}c@{}c@{}c@{}c@{}c@{}c@{}c@{}c@{}c@{}
                                c@{}c@{}c@{}c@{}c@{}c@{}c@{}c@{}c@{}c@{}
                                c@{}c@{}c@{}c@{}c@{}c@{}c@{}c@{}c@{}c@{}
                                c@{}c@{}c@{}c@{}c@{}
                                                            c@{}|}
%                                                            c@{\hspace{1.0mm}}|}
\hline
 & &  \multicolumn{36}{ c|}{\mbox{                  }} \\ 
 & &  \multicolumn{36}{ c|}{\normalsize \em (001000) } \\
 & &  \multicolumn{36}{ c|}{\mbox{                  }} \\ 
\cline{3-38}
 & & 
          \AA 1\ZZ& \AA 2\ZZ& \AA 3\ZZ& \AA 4\ZZ& \AA 5\ZZ& \AA 6\ZZ& \AA 7\ZZ& \AA 8\ZZ& \AA 9\ZZ& \AA 10\ZZ& 
 \AA 11\ZZ& \AA 12\ZZ& \AA 13\ZZ& \AA 14\ZZ& \AA 15\ZZ& \AA 16\ZZ& \AA 17\ZZ& \AA 18\ZZ& \AA 19\ZZ& \AA 20\ZZ&
 \AA 21\ZZ& \AA 22\ZZ& \AA 23\ZZ& \AA 24\ZZ& \AA 25\ZZ& \AA 26\ZZ& \AA 27\ZZ& \AA 28\ZZ& \AA 29\ZZ& \AA 30\ZZ&
 \AA 31\ZZ& \AA 32\ZZ& \AA 33\ZZ& \AA 34\ZZ& \AA 35\ZZ& \AA 36\ZZ \\
\hline
%                         &1  2  3  4  5  6  7  8  9  0* 1  2  3  4  5  6  7  8  9  0* 1  2  3  4  5  6  7  8  9  0* 1  2  3  4  5  6
%                          -------------------------------------------------------------------------------------------------------------
        &                 &  &  &  &  &  &  &  &  &  &  &  &  &  &  &  &  &  &  &  &  &  &  &  &  &  &  &  &  &  &  &  &  &  &  &  &  \\
\mbmm\NA&\NA              &  &  &  &  &  &  &  &  &  &  &  &  &  &  &  &  &  &  &  &  &  &  &  &  &  &  &  &  &  &  &  &  &  &  &  &  \\
\mbmm\NB&\NB              &  &  &  &  &  &  &  &  &  &  &  &  &  &  &  &  &  &  &  &  &  &  &  &  &  &  &  &  &  &  &  &  &  &  &  &  \\
\mbmm\NC&\NC              &  &  &  &  &  &  &  &  &  &  &  &  &  &  &  &  &  &  &  &  &  &  &  &  &  &  &  &  &  &  &  &  &  &  &  &  \\
\mbmm\ND&\ND              &  &  &  &  &  &  &  &  &  &  &  &  &  &  &  &  &  &  &  &  &  &  &  &  &  &  &  &  &  &  &  &  &  &  &  &  \\
\mbmm\NE&\NE              &  &  &  &  &  &  &  &  &  &  &  &  &  &  &  &  &  &  &  &  &  &  &  &  &  &  &  &  &  &  &  &  &  &  &  &  \\
\mbmm\NF&\NF              &  &  &  &  &  &  &  &  &  &  &  &  &  &  &  &  &  &  &  &  &  &  &  &  &  &  &  &  &  &  &  &  &  &  &  &  \\
        &                 &  &  &  &  &  &  &  &  &  &  &  &  &  &  &  &  &  &  &  &  &  &  &  &  &  &  &  &  &  &  &  &  &  &  &  &  \\
\mbmm\NA&\NB              &  &  &  &  &  &  &  &  &  &  &  &  &  &  &  &  &  &  &  &  &  &  &  &  &  &  &  &  &  & 2&-2&  & 2&  &  &  \\
\mbmm\NA&\NC              &  &  &  &  &  &  &  &  &  &  &  &  &  &  &  &  &  &  &  &  &  &  &  &  &  &  &  &  &  &  &  &  &  &  &  &  \\
\mbmm\NA&\ND              &  &  &  &  &  &  &  &  &  &  &  &  &  &  &  &  &  &  &  &  &  &  &  &  &  &  &  &  &  &  &  &  &  &  &  & 2\\
\mbmm\NA&\NE              &  &  &  &  &  &  &  &  &  &  &  &  &  &  &  &  &  &  &  & 2&-2&  &  &  &  &  &  &  &  &  &  &  &  &  &  &  \\
\mbmm\NA&\NF              &  &  &  &  &  &  &  &  &  &  &-2& 2&  &  &  &  &  &  &  &  &  &  &  &  &  &  &  &  &  &  &  &  &  &  &  &  \\
\mbmm\NB&\NC              &  &  &  &  &  &  &  &  &  &  &  &  &  &  &  &  &  &  &  &  &  &  &  &  &  &  &  &  &  &  &  &-2& 2&  &  &  \\
\mbmm\NB&\ND              &  &  &  &  &  &  &  &  &  &  &  &  &  &  &  &  &  &  &  &  &  &  &  &  &  &  &-2&  & 2&  &  &  &  &  &  &  \\
\mbmm\NB&\NE              &  &  &  &  &  &  &  &  &  &  &  &  &  &  &  &-2&  & 2&  &  &  &  &  &  &  &  &  &  &  &  &  &  &  &  &  &  \\
\mbmm\NB&\NF              &  &  &  &  &  &-2& 2&  &  &  &  &  &  &  &  &  &  &  &  &  &  &  &  &  &  &  &  &  &  &  &  &  &  &  &  &  \\
\mbmm\NC&\ND              &  &  &  &  &  &  &  &  &  &  &  &  &  &  &  &  &  &  &  &  &  &  &  &  &  &  &  &  &  &  &  &  &  &  &  &  \\
\mbmm\NC&\NE              &  &  &  &  &  &  &  &  &  &  &  &  &  &  &  &  &  &  &  &  &  &  &-2& 2&  &  &  &  &  &  &  &  &  &  &  &  \\
\mbmm\NC&\NF              &  &  &  &  &  &  &  &  &  &  &  &  &-2& 2&  &  &  &  &  &  &  &  &  &  &  &  &  &  &  &  &  &  &  &  &  &  \\
\mbmm\ND&\NE              &  &  &  &  &  &  &  &  &  &  &  &  &  &  &  &  &  &  &  &  &  &  &  &  &-2& 2&  &  &  &  &  &  &  &  &  &  \\
\mbmm\ND&\NF              &  &  &  &  &  &  &  &  &-2& 2&  &  &  &  &  &  &  &  &  &  &  &  &  &  &  &  &  &  &  &  &  &  &  &  &  &  \\
\mbmm\NE&\NF              &  &  &-2& 2&  &  &  &  &  &  &  &  &  &  &  &  &  &  &  &  &  &  &  &  &  &  &  &  &  &  &  &  &  &  &  &  \\
\AA000001\ZA00000\1\ZZ    &  &  &  &  &  &  &  &  &  &  &  &  &  &  &-1&  &  &  &  &  &  &  &  &  &  &  &  &  &  &  &  &  &  &  &  &  \\
\AA00100\1\ZA00\1001\ZZ   &  &  &  &  &  &  &  &  &  &  &  &  &  &  &-1&  &  &  &  &  &  &  &  &  &  &  &  &  &  &  &  &  &  &  &  &  \\
\AA01\1100\ZA0\11\100\ZZ  &  &  &  &  &  &  &  &  &  &  &  &  &  &  &  &  &  &  &  &  &  &  &  &  &  &  &  &-1&  &  &  &  &  &  &  &  \\
\AA1\10100\ZA\110\100\ZZ  &  &  &  &  &  &  &  &  &  &  &  &  &  &  &  &  &  &  &  &  &  &  &  &  &  &  &  &-1&  &  &  &  &  &  &-1&  \\
\AA010\110\ZA0\101\10\ZZ  &  &  &  &  &  &  &  &  &  &  &  &  &  &  &  &  &-1&  &  &  &  &  &  &  &  &  &  &-1&  &  &  &  &  &  &  &  \\ 
\AA\100100\ZA100\101\ZZ   &  &  &  &  &  &  &  &  &  &  &  &  &  &  &  &  &  &  &  &  &  &  &  &  &  &  &  &  &  &  &  &  &  &  &-1&  \\ 
\AA1\11\110\ZA\11\11\11\ZZ&  &  &  &  &  &  &  &  &  &  &  &  &  &  &  &  &-1&  &-1&  &  &  &  &  &  &  &  &-1&  &  &  &  &  &  &-1&  \\ 
\AA0100\10\ZA0\10011\ZZ   &  &  &  &  &  &  &  &  &  &  &  &  &  &  &  &  &-1&  &  &  &  &  &  &  &  &  &  &  &  &  &  &  &  &-1&  &  \\ 
%5:
\AA\101\110\ZA10\11\11\ZZ &  &  &  &  &  &  &  &  &  &  &  &  &  &  &  &  &  &  &-1&  &  &-1&  &  &  &  &  &  &  &  &  &  &  &  &-1&  \\
\AA10\1011\ZA\1010\10\ZZ  &  &-1&  &  &  &  &  &  &  &  &  &  &  &  &-1&  &  &  &-1&  &  &  &  &  &  &  &  &  &  &  &  &  &  &  &  &  \\
\AA1\110\10\ZA\11\1011\ZZ &  &  &  &  &  &  &  &  &  &  &  &  &  &  &  &  &-1&  &-1&  &  &  &  &  &  &  &  &  &  &  &  &  &  &-1&  &  \\
%6:
\AA\1010\10\ZA10\1010\ZZ  &  &  &  &  &  &  &  &  &  &  &  &  &  &  &  &  &  &  &-1&  &  &-1&  &  &  &  &  &  &  &  &  &  &  &  &  &  \\
\AA\11\1011\ZA1\110\1\1\ZZ&-1&-1&  &  &  &  &  &  &  &  &  &  &  &  &  &  &-1&  &-1&  &  &-1&  &  &  &  &  &  &  &  &  &  &  &  &  &  \\
\AA10001\1\ZA\1000\11\ZZ  &  &-1&  &  &  &  &  &  &  &  &  &  &  &  &-1&  &  &  &  &-1&-1&  &  &  &  &  &  &  &  &  &  &  &  &  &  &  \\ 
\AA10\11\11\ZA\101\11\1\ZZ&  &-1&  &  &  &  &  &-1&  &  &  &  &  &  &  &  &  &  &-1&  &  &  &  &  &  &  &  &  &  &  &  &  &  &  &-1&  \\ 
%7:
\AA\11\11\11\ZA1\11\11\1\ZZ&-1&-1& &  &-1&  &  &-1&  &  &  &  &  &  &-1&  &-1&  &-1&  &  &-1&  &  &  &  &  &-1&  &  &  &  &  &-1&-1&  \\
\AA0\10011\ZA0100\1\1\ZZ  &-1&  &-1&-1&  &  &  &  &  &  &  &  &  &  &  &  &-1&  &  &  &  &  &  &  &  &  &  &  &  &  &  &  &  &  &  &  \\
\AA\11001\1\ZA1\100\11\ZZ &-1&-1&  &  &  &  &  &  &  &  &  &  &  &  &  &-1&  &-1&  &-1&-1&  &  &  &  &  &  &  &  &  &  &  &  &  &  &  \\
\AA1001\1\1\ZA\100\111\ZZ &  &-1&  &  &  &  &  &-1&  &  &  &  &  &  &  &  &  &  &  &-1&-1&  &  &  &  &  &  &  &  &  &  &  &  &  &  &-1\\
\AA100\101\ZA\10010\1\ZZ  &  &  &  &  &  &  &  &-1&  &  &-1&-1&  &  &  &  &  &  &  &  &  &  &  &  &  &  &  &  &  &  &  &  &  &  &-1&  \\
%8:
\AA0\101\11\ZA010\11\1\ZZ &-1&  &-1&-1&-1&  &  &  &-1&-1&  &  &  &  &  &  &-1&  &  &  &  &  &  &  &  &  &  &-1&  &  &  &  &  &-1&  &  \\
\AA\110\101\ZA1\1010\1\ZZ &  &  &  &  &-1&-1&-1&-1&  &  &-1&-1&  &  &  &  &  &  &  &  &  &  &  &  &  &  &  &-1&  &  &  &  &  &  &-1&  \\
\AA\1101\1\1\ZA1\10\111\ZZ&-1&-1&  &  &-1&  &  &-1&  &  &  &  &  &  &-1&-1&  &-1&  &-1&-1&  &  &  &  &  &-1&  &-1&  &  &  &  &  &  &-1\\
\AA0\1101\1\ZA01\10\11\ZZ &-1&  &-1&-1&  &  &  &  &  &  &  &  &  &  &  &-1&  &-1&  &  &  &-1&-1&-1&  &  &  &  &  &  &  &  &  &  &  &  \\
\AA101\10\1\ZA\10\1101\ZZ &  &  &  &  &  &  &  &-1&  &  &-1&-1&  &  &  &  &  &  &  &  &  &  &  &  &  &  &  &  &  &  &  &  &  &  &  &-1\\
%9:
\AA0\11\101\ZA01\110\1\ZZ &  &  &  &  &-1&-1&-1&  &-1&-1&  &  &-1&-1&-1&  &  &  &  &  &  &  &  &  &  &  &  &-1&  &  &  &  &  &  &  &  \\
\AA0\111\1\1\ZA01\1\111\ZZ&-1&  &-1&-1&-1&  &  &  &-1&-1&  &  &  &  &  &-1&  &-1&  &  &  &-1&-1&-1&  &  &-1&  &-1&  &  &  &  &  &  &  \\
\AA\111\10\1\ZA1\1\1101\ZZ&  &  &  &  &-1&-1&-1&-1&  &  &-1&-1&  &  &  &  &  &  &  &  &  &  &  &  &  &  &-1&  &-1&  &  &-1&-1&-1&  &-1\\
\AA00\1110\ZA001\1\10\ZZ  &  &  &  &  &  &  &  &  &  &  &  &  &  &  &  &  &  &  &  &  &  &-1&-1&-1&-1&-1&  &  &  &  &  &  &  &  &  &  \\
\AA11\1000\ZA\1\11000\ZZ  &  &  &  &  &  &  &  &  &  &  &  &  &  &  &  &  &  &  &  &  &  &  &  &  &  &  &  &  &  &-1&-1&  &-1&-1&  &  \\
%10:
\AA2\10000\ZA\210000\ZZ   &  &  &  &  &  &  &  &  &  &  &  &  &  &  &  &  &  &  &  &  &  &  &  &  &  &  &  &  &  &-1&-1&  &-1&-1&  &  \\
\AA\12\1000\ZA1\21000\ZZ  &  &  &  &  &  &  &  &  &  &  &  &  &  &  &  &  &  &  &  &  &  &  &  &  &  &  &  &  &  &-1&-1&-1&-2&-2&  &  \\
\AA0\12\10\1\ZA01\2101\ZZ &  &  &  &  &-1&-1&-1&  &-1&-1&  &  &-1&-1&-1&  &  &  &  &  &  &  &  &  &  &  &-1&  &-1&  &  &-1&-1&-1&  &  \\
\AA00\12\10\ZA001\210\ZZ  &  &  &  &  &  &  &  &  &  &  &  &  &  &  &  &  &  &  &  &  &  &-1&-1&-1&-1&-1&  &  &  &  &  &  &  &  &  &  \\
\AA000\120\ZA0001\20\ZZ   &  &  &  &  &  &  &  &  &  &  &  &  &  &  &  &  &  &  &  &  &  &  &  &  &-1&-1&  &  &  &  &  &  &  &  &  &  \\
\AA00\1002\ZA00100\2\ZZ   &  &  &  &  &  &  &  &  &  &  &  &  &-1&-1&-2&  &  &  &  &  &  &  &  &  &  &  &  &  &  &  &  &  &  &  &  &  \\
        &                 &  &  &  &  &  &  &  &  &  &  &  &  &  &  &  &  &  &  &  &  &  &  &  &  &  &  &  &  &  &  &  &  &  &  &  &  \\
%                          -------------------------------------------------------------------------------------------------------------
%                          &1  2  3  4  5  6  7  8  9  0* 1  2  3  4  5  6  7  8  9  0* 1  2  3  4  5  6  7  8  9  0* 1  2  3  4  5  6
\hline			    
        &                 &  &  &  &  &  &  &  &  &  &  &  &  &  &  &  &  &  &  &  &  &  &  &  &  &  &  &  &  &  &  &  &  &  &  &  &  \\
  & &           4  &         4  &        4   &        4   &        4   &         4  &   4  &   4  &   4  &    4  &
                4  &         4  & $\sqrt{12}$& $\sqrt{12}$& $\sqrt{24}$&         4  &   4  &   4  &   4  &    4  &
                4  &         4  &        4   &        4   & $\sqrt{12}$& $\sqrt{12}$&   4  &   4  &   4  & $\sqrt{12}$&
        $\sqrt{12}$& $\sqrt{12}$& $\sqrt{32}$& $\sqrt{24}$&        4   &         4 \\
        &                 &  &  &  &  &  &  &  &  &  &  &  &  &  &  &  &  &  &  &  &  &  &  &  &  &  &  &  &  &  &  &  &  &  &  &  &  \\
\hline
\end{tabular}
\end{table}

\clearpage

\protect
\begin{table}[t]
\caption{ 
{\bf CG coefficients for the (000000) dominant weight states of the
\hspace*{8.0in}.   
         2925-dimensional (001000) irrep,  650-dimensional (100010) irrep,   
           78-dimensional 
\hspace*{8.0in}.  
                          (000001) irrep, and the singlet {\em S}\/
         in the product (000001)$\otimes$(000001).}
\AA n\ZZ\ $\equiv$ \AA000000$_n$\ZZ. 
\hspace*{8.0in}.  
Each CGC should be divided by the respective number in the last row to maintain
$\langle n\!\mid\! n\rangle = 1$.
\hspace*{8.0in}.  
}
\label{t:cgc000000_2925_650_78_S}
\tiny
%                            *         *         *         
%                 12  345678901234567890123456789012345678
%\begin{tabular}{|cc||rrrrrrrrrrrrrrrrrrrrrrrrrrrrrrrrrrrr|}
\begin{tabular}{|@{\hspace{0.5mm}}l@{\hspace{0.5mm}}l@{\hspace{1mm}}
                |@{\hspace{0mm}}c@{}c@{}c@{}c@{}c@{}c@{}c@{}c@{}c@{$\,$}|@{}
                                r@{}r@{}r@{}r@{}r@{}r@{}r@{}r@{}r@{}r@{}
                                r@{}r@{}r@{}r@{}r@{}r@{}r@{}r@{}r@{}r@{$\,$}|@{}
                                r@{}r@{}r@{}r@{}r@{}r@{$\,$}|@{}
%                                                            r@{}|}
                                                            r@{\hspace{1.0mm}}|}
\hline
 & &  \multicolumn{ 9}{ c|}{\mbox{ }}  &  \multicolumn{20}{ c|}{\mbox{ }}  &  \multicolumn{6}{ c|}{\mbox{ }}  &  \\
 & &  \multicolumn{ 9}{ c|}{\normalsize \em (001000) }  &
      \multicolumn{20}{ c|}{\normalsize \em (100010) }  &
      \multicolumn{ 6}{ @{$\,$}c@{$\,$}|}{\normalsize \em (000001) }  & {\normalsize \em S }  \\ 
 & &  \multicolumn{ 9}{ c|}{continued from table \ref{t:cgc000000_2925}}  &  
      \multicolumn{20}{ c|}{\mbox{ }}  &  \multicolumn{6}{ c|}{\mbox{ }}  &  \\
 & &  \multicolumn{ 9}{ c|}{\mbox{ }}  &  \multicolumn{20}{ c|}{\mbox{ }}  &  \multicolumn{6}{ c|}{\mbox{ }}  &  \\
\cline{3-38}
 & & 
            \AA 37\ZZ& \AA 38\ZZ& \AA 39\ZZ& \AA 40\ZZ& \AA 41\ZZ& \AA 42\ZZ& \AA 43\ZZ& \AA 44\ZZ& \AA 45\ZZ&
          \AA 1\ZZ& \AA 2\ZZ& \AA 3\ZZ& \AA 4\ZZ& \AA 5\ZZ& \AA 6\ZZ& \AA 7\ZZ& \AA 8\ZZ& \AA 9\ZZ& \AA 10\ZZ& 
 \AA 11\ZZ& \AA 12\ZZ& \AA 13\ZZ& \AA 14\ZZ& \AA 15\ZZ& \AA 16\ZZ& \AA 17\ZZ& \AA 18\ZZ& \AA 19\ZZ& \AA 20\ZZ&
          \AA 1\ZZ& \AA 2\ZZ& \AA 3\ZZ& \AA 4\ZZ& \AA 5\ZZ& \AA 6\ZZ& 
          \AA 1\ZZ \\
\hline
%                         &7  8  9  0  1  2  3  4  5*  1   2   3   4   5   6   7   8   9   0*  1   2   3   4   5   6   7   8   9   0*  1   2   3   4   5   6*   1
%                          ---------------------------------------------------------------------------------------------------------------------------------------
        &                 &  &  &  &  &  &  &  &  &  &   &   &   &   &   &   &   &   &   &   &   &   &   &   &   &   &   &   &   &   &   &   &   &   &   &   &   \\
\mbmm\NA&\NA              &  &  &  &  &  &  &  &  &  &\pD&   &   &   &   &   &   &   &   &   &   &   &   &   &   &   &   &   &   &   &   &   &   &   &   &   &  8\\
\mbmm\NB&\NB              &  &  &  &  &  &  &  &  &  &   &   &\mD&\pD&   &   &   &   &   &   &   &   &   &   &   &   &   &   &   &   &   &   &   &   &   &   & 20\\
\mbmm\NC&\NC              &  &  &  &  &  &  &  &  &  &   &   &   &   &   &\pD&\pD&\pD&   &   &   &   &   &   &   &   &   &   &   &   &   &   &   &   &   &   & 36\\
\mbmm\ND&\ND              &  &  &  &  &  &  &  &  &  &   &   &   &   &   &   &   &   &   &\pD&\mD&   &   &   &   &   &   &   &   &   &   &   &   &   &   &   & 20\\
\mbmm\NE&\NE              &  &  &  &  &  &  &  &  &  &   &   &   &   &   &   &   &   &   &   &   &   &   &   &\pD&   &   &   &   &   &   &   &   &   &   &   &  8\\
\mbmm\NF&\NF              &  &  &  &  &  &  &  &  &  &   &   &   &   &   &   &   &   &   &   &   &   &   &   &   &   &   &\mD&   &   &   &   &   &   &   &   & 12\\
        &                 &  &  &  &  &  &  &  &  &  &   &   &   &   &   &   &   &   &   &   &   &   &   &   &   &   &   &   &   &   &   &   &   &   &   &   &   \\
\mbmm\NA&\NB              &  &  &  &  &  &  &  &  &  &\mD&   &   &\mB&   &   &   &   &   &   &   &   &   &   &   &   &   &   &   &   &   &   &   &   &   &   &-10\\
\mbmm\NA&\NC              &  &  &  &  &  & 2&-2&  &  &\pD&   &   &   &   &   &   &   &\mB&   &   &   &   &   &   &   &   &   &   &   &   &   &   &   &   &   & 12\\
\mbmm\NA&\ND              &-2&  &  &  &  &  &  &  &  &\mB&\mB&   &   &   &   &   &   &   &   &   &   &   &   &   &   &   &   &   &   &   &   &   &   &   &   & -8\\
\mbmm\NA&\NE              &  &  &  &  &  &  &  &  &  &   &   &   &   &   &   &   &   &   &   &   &\mB&   &   &   &   &   &   &   &   &   &   &   &   &   &   &  4\\
\mbmm\NA&\NF              &  &  &  &  &  &  &  &  &  &\mB&   &   &   &   &   &   &   &   &   &   &   &   &   &   &   &   &   &\mB&   &   &   &   &   &   &   & -6\\
\mbmm\NB&\NC              &  &  &  &  &  &  &  &  &  &   &   &\pD&\mB&   &   &\mB&\mB&   &   &   &   &   &   &   &   &   &   &   &   &   &   &   &   &   &   &-24\\
\mbmm\NB&\ND              &  &  &  &  &  &  &  &  &  &   &   &\mB&   &\mB&   &   &   &   &   &\mB&   &   &   &   &   &   &   &   &   &   &   &   &   &   &   & 16\\
\mbmm\NB&\NE              &  &  &  &  &  &  &  &  &  &   &   &   &   &   &   &   &   &   &   &   &   &\mB&   &\mB&   &   &   &   &   &   &   &   &   &   &   & -8\\
\mbmm\NB&\NF              &  &  &  &  &  &  &  &  &  &   &   &\mB&   &   &   &   &   &   &   &   &   &   &   &   &   &   &\mB&   &\mB&   &   &   &   &   &   & 12\\
\mbmm\NC&\ND              &  & 2&-2&  &  &  &  &  &  &   &   &   &   &   &\mB&   &\mB&   &\mB&\pD&   &   &   &   &   &   &   &   &   &   &   &   &   &   &   &-24\\
\mbmm\NC&\NE              &  &  &  &  &  &  &  &  &  &   &   &   &   &   &   &   &   &   &   &   &   &   &\mB&\pD&   &   &   &   &   &   &   &   &   &   &   & 12\\
\mbmm\NC&\NF              &  &  &  &  &  &  &  &  &-2&   &   &   &   &   &\mB&\mB&   &   &   &   &   &   &   &   &   &   &\pD&   &   &   &   &   &   &   &   &-18\\
\mbmm\ND&\NE              &  &  &-2&  &  &  &  &  &  &   &   &   &   &   &   &   &   &   &\mB&   &   &   &   &\mD&   &   &   &   &   &   &   &   &   &   &   &-10\\
\mbmm\ND&\NF              &  &  &  &  &  &  &  &  &  &   &   &   &   &   &   &   &   &   &   &\mB&   &   &   &   &\mB&   &\mB&   &   &   &   &   &   &   &   & 12\\
\mbmm\NE&\NF              &  &  &  &  &  &  &  &  &  &   &   &   &   &   &   &   &   &   &   &   &   &   &   &\mB&   &\mB&   &   &   &   &   &   &   &   &   & -6\\
\AA000001\ZA00000\1\ZZ    &  &  &  &  &  &  &  &  &  &   &   &   &   &   &   &   &   &   &   &   &   &   &   &   &   &   &\p1&   &   &   &   &   &   &   &\p1& -3\\
\AA00100\1\ZA00\1001\ZZ   &  &  &  &  &  &  &  &  &-1&   &   &   &   &   &   &   &\p1&   &   &   &   &   &   &   &   &   &\p1&   &   &   &   &\p1&   &   &\p1&  3\\
\AA01\1100\ZA0\11\100\ZZ  &  &  &  &  &  &  &  &  &-1&   &   &   &   &\p1&   &   &\p1&   &   &   &   &   &   &   &   &   &   &   &   &   &\p1&\p1&\p1&   &   & -3\\
\AA1\10100\ZA\110\100\ZZ  &  &  &  &  &  &  &  &  &  &   &\p1&   &   &\p1&   &   &   &   &\p1&   &   &   &   &   &   &   &   &   &   &\p1&\p1&   &\m1&   &   &  3\\
\AA010\110\ZA0\101\10\ZZ  &  &  &  &  &  &  &  &  &  &   &   &   &\p1&\p1&   &   &   &   &   &   &   &\p1&   &   &   &   &   &   &   &   &\m1&   &\p1&\p1&   &  3\\
\AA\100100\ZA100\101\ZZ   &  &  &  &-1&  &  &  &  &  &   &\p1&   &   &   &   &   &   &   &\m1&\p1&   &   &   &   &   &   &   &   &   &\p1&   &   &\p1&   &   & -3\\
\AA1\11\110\ZA\11\11\11\ZZ&  &  &  &  &  &  &  &  &-1&   &\p1&   &\p1&\p1&   &   &\p1&\p1&\p1&   &\p1&\p1&\p1&   &   &   &   &   &   &\m1&\m1&\m1&\m1&\m1&   & -3\\
\AA0100\10\ZA0\10011\ZZ   &  &  &  &  &  &  &  &  &  &   &   &\p1&\m1&   &   &   &   &   &   &   &   &\p1&   &   &   &   &   &   &   &   &\p1&   &\m1&\m1&   & -3\\
%5:
\AA\101\110\ZA10\11\11\ZZ &  &  &  &-1&  &  &  &  &  &   &\p1&   &   &   &   &\p1&   &\p1&\m1&\p1&\p1&   &\m1&\p1&   &   &   &   &   &\m1&   &\p1&\p1&\p1&   &  3\\
\AA10\1011\ZA\1010\10\ZZ  &  &  &  &  &  &  &  &  &-1&   &   &   &   &   &   &   &\p1&\p1&   &   &\m1&   &\p1&   &   &\p1&\p1&\p1&   &\p1&   &\m1&   &\p1&\m1&  3\\
\AA1\110\10\ZA\11\1011\ZZ &  &  &  &  &-1&  &  &  &  &\p1&   &\p1&\p1&   &\p1&   &   &\m1&   &   &\p1&\p1&\p1&   &   &   &   &   &   &\p1&\p1&\p1&   &\m1&   &  3\\
%6: 
\AA\1010\10\ZA10\1010\ZZ  &  &  &  &  &-1&  &  &-1&-1&\p1&   &   &   &   &\m1&\m1&\m1&\m1&   &   &\p1&   &\m1&\p1&   &   &   &   &   &\p1&   &\m1&   &\p1&   & -3\\
\AA\11\1011\ZA1\110\1\1\ZZ&  &  &  &  &  &  &  &  &  &   &   &   &\p1&   &   &\p1&   &\p1&   &   &\m1&\m1&\m1&\m1&   &\m1&   &\p1&\p1&\p1&\p1&\p1&   &\m1&\p1& -3\\
\AA10001\1\ZA\1000\11\ZZ  &  &  &  &  &  &  &  &  &  &   &   &   &   &   &   &   &   &   &   &   &\p1&   &   &   &   &\p1&\p1&\p1&   &\m1&   &   &   &\m1&\m1& -3\\
\AA10\11\11\ZA\101\11\1\ZZ&  &  &  &  &-1&  &  &  &  &\m1&\m1&   &   &   &\p1&   &   &\m1&\p1&   &\m1&   &\p1&   &\p1&\p1&   &\m1&   &\m1&   &\p1&\p1&\p1&\p1& -3\\
%7: 
\AA\11\11\11\ZA1\11\11\1\ZZ& &  &  &-1&-1&  &  &-1&-1&\m1&\m1&\m1&\m1&\m1&\m1&\m1&\m1&\m1&\m1&\m1&\m1&\m1&\m1&\m1&\m1&\m1&\m1&\m1&\m1&\m1&\m1&\m1&\m1&\m1&\m1&  3\\
\AA0\10011\ZA0100\1\1\ZZ  &  &  &  &  &  &  &  &  &  &   &   &   &\p1&   &   &   &   &   &   &   &   &\m1&   &\p1&   &\p1&   &   &\p1&   &\p1&   &   &\p1&\m1&  3\\
\AA\11001\1\ZA1\100\11\ZZ &  &  &  &  &  &  &  &  &  &   &   &   &   &   &   &   &   &   &   &   &\p1&\p1&   &\p1&   &\m1&   &\p1&\p1&\m1&\m1&   &   &\p1&\p1&  3\\
\AA1001\1\1\ZA\100\111\ZZ &-1&  &  &  &  &  &  &  &  &\p1&\p1&   &   &   &   &   &   &   &   &   &\p1&   &   &   &\p1&\p1&   &\m1&   &\p1&   &   &\m1&\m1&\p1&  3\\
\AA100\101\ZA\10010\1\ZZ  &  &  &  &  &  &  &  &  &  &\p1&\m1&   &   &   &   &   &   &   &\p1&   &   &   &   &   &\p1&   &   &\p1&   &\p1&   &   &\p1&   &\m1&  3\\
%8:
\AA0\101\11\ZA010\11\1\ZZ &  &  &  &  &  &  &  &  &  &   &   &\m1&\m1&\m1&   &   &   &   &   &\p1&   &\m1&   &\p1&\p1&\p1&\p1&   &\m1&   &\m1&   &\p1&\p1&\p1& -3\\
\AA\110\101\ZA1\1010\1\ZZ &  &  &  &-1&  &  &  &  &  &\p1&\m1&\p1&   &\m1&   &   &   &   &\m1&\m1&   &   &   &   &\m1&   &\p1&\p1&\p1&\p1&\p1&   &\m1&   &\p1& -3\\
\AA\1101\1\1\ZA1\10\111\ZZ&-1&  &  &  &  &  &  &  &  &\p1&\p1&\p1&   &\p1&   &   &   &   &   &\p1&\p1&\p1&   &\p1&\m1&\m1&\m1&\m1&\m1&\p1&\p1&   &\p1&\p1&\m1& -3\\
\AA0\1101\1\ZA01\10\11\ZZ &  &  &  &  &  &  &  &  &  &   &   &   &   &   &   &\p1&   &   &   &   &   &\p1&\p1&\m1&   &\p1&   &   &\p1&   &   &   &   &   &   & -3\\
\AA101\10\1\ZA\10\1101\ZZ &-1&  &  &  &-1&-1&-1&  &  &\m1&\p1&   &   &   &\p1&   &   &\p1&   &   &   &   &   &   &\p1&   &   &\p1&   &\m1&   &\m1&\m1&   &\m1& -3\\
%9:
\AA0\11\101\ZA01\110\1\ZZ &  &  &  &  &  &  &  &-1&-2&   &   &\p1&   &\m1&\p1&\p1&\m1&   &   &\p1&   &   &   &   &\p1&   &\m1&   &\p1&   &\p1&\p1&\p1&   &\m1&  3\\
\AA0\111\1\1\ZA01\1\111\ZZ&  &-1&-1&-1&  &  &  &-1&  &   &   &\p1&   &\p1&\p1&\m1&\p1&   &\p1&\m1&   &\p1&\p1&\m1&\p1&\p1&\p1&   &\m1&   &\p1&\p1&\m1&\m1&\p1&  3\\
\AA\111\10\1\ZA1\1\1101\ZZ&-1&  &  &  &-1&-1&-1&-1&  &\m1&\p1&\m1&\p1&\p1&\m1&\p1&\p1&\p1&   &\p1&   &   &   &   &\m1&   &\p1&\p1&\p1&\m1&\m1&\p1&\p1&   &\p1&  3\\
\AA00\1110\ZA001\1\10\ZZ  &  &  &-1&-1&  &  &  &  &  &   &   &   &   &   &   &\p1&   &   &\p1&\p1&   &   &\p1&\p1&   &   &   &   &   &   &   &\m1&\m1&\p1&   &  3\\
\AA11\1000\ZA\1\11000\ZZ  &  &  &  &  &-1&-1&-1&  &  &\p1&   &\p1&\p1&   &\p1&   &   &\p1&   &   &   &   &   &   &   &   &   &   &   &\p1&\m1&\m1&   &   &   &  3\\
%10:
\AA2\10000\ZA\210000\ZZ   &  &  &  &  &  &  &  &  &  &   &   &\p1&\p1&   &   &   &   &   &   &   &   &   &   &   &   &   &   &   &   &\mB&\m1&   &   &   &   & -3\\
\AA\12\1000\ZA1\21000\ZZ  &  &  &  &  &-1&-1&-1&-1&  &\p1&   &   &   &   &\m1&\p1&\p1&\p1&   &   &   &   &   &   &   &   &   &   &   &\p1&\pB&\p1&   &   &   & -3\\
\AA0\12\10\1\ZA01\2101\ZZ &  &-1&-1&-1&  &  &  &-2&-1&   &   &\m1&\p1&\p1&   &   &   &   &\p1&\m1&   &   &   &   &\p1&   &\m1&   &\p1&   &\m1&\mB&\m1&   &\m1& -3\\
\AA00\12\10\ZA001\210\ZZ  &  &-1&-2&-2&  &  &  &-1&  &   &   &   &   &   &\p1&\m1&\p1&   &   &   &   &   &\p1&\p1&   &   &   &   &   &   &   &\p1&\pB&\p1&   & -3\\
\AA000\120\ZA0001\20\ZZ   &  &  &-1&-1&  &  &  &  &  &   &   &   &   &   &   &   &   &   &\p1&\p1&   &   &   &   &   &   &   &   &   &   &   &   &\m1&\mB&   & -3\\
\AA00\1002\ZA00100\2\ZZ   &  &  &  &  &  &  &  &-1&-2&   &   &   &   &   &\p1&\p1&\m1&   &   &   &   &   &   &   &   &   &   &   &   &   &   &\p1&   &   &\pB& -3\\
        &                 &  &  &  &  &  &  &  &  &  &   &   &   &   &   &   &   &   &   &   &   &   &   &   &   &   &   &   &   &   &   &   &   &   &   &   &   \\
%                          --------------------------------------------------------------------------------------------------------------------------------------
%                         &7  8  9  0  1  2  3  4  5*  1   2   3   4   5   6   7   8   9   0*  1   2   3   4   5   6   7   8   9   0*  1   2   3   4   5   6*   1
\hline			    
  & & & & & & & & & & & 
         \multicolumn{20}{c|}{   } &
         \multicolumn{ 6}{@{$\,$}c@{$\,$}|}{ } & \\
  & &           4  & $\sqrt{12}$& $\sqrt{32}$& $\sqrt{24}$&        4   &         4  &         4  & $\sqrt{26}$&         6  &
         \multicolumn{20}{c|}{$\sqrt{32}$ each state} &
         \multicolumn{ 6}{@{$\,$}c@{$\,$}|}{$\sqrt{48}$ each state} &
        $\sqrt{702}$ \\
  & & & & & & & & & & & 
         \multicolumn{20}{c|}{   } &
         \multicolumn{ 6}{@{$\,$}c@{$\,$}|}{ } & \\
\hline
\end{tabular}
\end{table}

\clearpage

\protect
\begin{table}[t]
\caption{ 
{\bf Embeddings of the $SU(3)_c\otimes SU(2)_L$ states into the {\bf 27} in $E_6$. }
\vspace*{2ex}
Signs follow from our choice of projections, eqs.(\ref{eq:proj_SO10}-\ref{eq:proj_SU3xSU2}).
\vspace*{2ex}  
}
\label{t:27states}
\tiny
\begin{tabular}{|c|rc|rc|rc|cr|}
\hline
                    &\multicolumn{2}{c|}{ } &\multicolumn{2}{c|}{ } &\multicolumn{2}{c|}{ } &\multicolumn{2}{c|}{ }\\
                    &\multicolumn{2}{c|}{ } &\multicolumn{2}{c|}{ } &\multicolumn{2}{c|}{ } &\multicolumn{2}{c|}{ }\\
{\small Superfield} & 
                                 \multicolumn{2}{c|}{\small SU(3)$_c\otimes$SU(2)$_L$ } &
                                 \multicolumn{2}{c|}{\small SU(5)} &
                                 \multicolumn{2}{c|}{\small SO(10)} &
                                 \multicolumn{2}{c|}{\small E$_6$} \\
                    &\multicolumn{2}{c|}{ } &\multicolumn{2}{c|}{ } &\multicolumn{2}{c|}{ } &\multicolumn{2}{c|}{ }\\
                    &\multicolumn{2}{c|}{ } &\multicolumn{2}{c|}{ } &\multicolumn{2}{c|}{ } &\multicolumn{2}{c|}{ }\\
standard &        &         &        &       &        &       &\multicolumn{2}{c|}{(100000) irrep}\\ 
embedding& weight &  irrep  & weight & irrep & weight & irrep & weight & level \\
                  &\multicolumn{2}{c|}{ } &\multicolumn{2}{c|}{ } &\multicolumn{2}{c|}{ } &\multicolumn{2}{c|}{ }\\
\hline
      &           &          &          &        &           &         &            &    \\
 Q    &  (10)(1)  &  (10)(1) &  (0100)  & (0100) &  (00001)  & (00001) &  (100000)  &  0 \\
      &           &          &          &        &           &         &            &    \\
      &-(\11)(1)  &          & (\1010)  &        &-(\10010)  &         & (1\10010)  &  7 \\
      &           &          &          &        &           &         &            &    \\
      &-(0\1)(1)  &          &-(\110\1) &        &-(0\1001)  &         & (10000\1)  & 11 \\
      &           &          &          &        &           &         &            &    \\
      &  (10)(\1) &          &-(10\11)  &        &-(010\10)  &         & (0000\11)  &  5 \\
      &           &          &          &        &           &         &            &    \\
      & (\11)(\1) &          &-(0\101)  &        &-(\1100\1) &         & (0\10001)  & 12 \\
      &           &          &          &        &           &         &            &    \\
      &-(0\1)(\1) &          & (00\10)  &        & (000\10)  &         & (0000\10)  & 16 \\
      &           &          &          &        &           &         &            &    \\
U$^c$ &  (01)(0)  &  (01)(0) & (1\110)  &        & (01\110)  &         & (00\1101)  &  3 \\
      &           &          &          &        &           &         &            &    \\
      & (1\1)(0)  &          &-(100\1)  &        &-(10\101)  &         & (01\11\10) &  7 \\
      &           &          &          &        &           &         &            &    \\
      & (\10)(0)  &          &-(0\11\1) &        &-(00\110)  &         & (00\1100)  & 14 \\
      &           &          &          &        &           &         &            &    \\
E$^c$ &  (00)(0)  &  (00)(0) & (\11\11) &        &-(\101\10) &         & (1\11\100) &  9 \\
      &           &          &          &        &           &         &            &    \\
D$^c$ &  (01)(0)  &  (01)(0) &  (0001)  & (0001) & (0010\1)  &         & (0\11000)  &  2 \\
      &           &          &          &        &           &         &            &    \\
      & (1\1)(0)  &          & (01\10)  &        &-(1\11\10) &         & (0010\1\1) &  6 \\
      &           &          &          &        &           &         &            &    \\
      & (\10)(0)  &          &-(\1000)  &        &-(0\110\1) &         & (0\1100\1) & 13 \\
      &           &          &          &        &           &         &            &    \\
 L    &-(00)(1)   &  (00)(1) &-(001\1)  &        & (1\1010)  &         & (00010\1)  &  4 \\
      &           &          &          &        &           &         &            &    \\
      & (00)(\1)  &          & (1\100)  &        & (1000\1)  &         & (\1001\10) &  9 \\
      &           &          &          &        &           &         &            &    \\
N$^c$ &-(00)(0)   &  (00)(0) &-(0000)   & (0000) &-(\11\101) &         & (10\1001)  & 10 \\
      &           &          &          &        &           &         &            &    \\
 T    &  (10)(0)  &  (10)(0) &  (1000)  & (1000) &  (10000)  & (10000) & (\110000)  &  1 \\
      &           &          &          &        &           &         &            &    \\
      &-(\11)(0)  &          &-(0\110)  &        & (0001\1)  &         & (\100010)  &  8 \\
      &           &          &          &        &           &         &            &    \\
      &-(0\1)(0)  &          & (000\1)  &        & (1\1000)  &         & (\11000\1) & 12 \\
      &           &          &          &        &           &         &            &    \\
H$_u$ &-(00)(1)   &  (00)(1) &-(\1100)  &        &-(0\1100)  &         & (001\11\1) &  5 \\
      &           &          &          &        &           &         &            &    \\
      & (00)(\1)  &          &-(00\11)  &        & (001\1\1) &         & (\101\100) & 10 \\
      &           &          &          &        &           &         &            &    \\
T$^c$ &-(01)(0)   &  (01)(0) &-(0001)   & (0001) &-(\11000)  &         & (000\111)  &  4 \\
      &           &          &          &        &           &         &            &    \\
      &-(1\1)(0)  &          &-(01\10)  &        &-(000\11)  &         & (010\100)  &  8 \\
      &           &          &          &        &           &         &            &    \\
      &-(\10)(0)  &          & (\1000)  &        &-(\10000)  &         & (000\110)  & 12 \\
      &           &          &          &        &           &         &            &    \\
H$_d$ & -(00)(1)  &  (00)(1) &-(001\1)  &        &-(00\111)  &         & (01\1010)  &  6 \\
      &           &          &          &        &           &         &            &    \\
      & (00)(\1)  &          & (1\100)  &        & (01\100)  &         & (\11\1001) & 11 \\
      &           &          &          &        &           &         &            &    \\
 S    &  (00)(0)  &  (00)(0) &  (0000)  & (0000) &  (00000)  & (00000) & (1\101\10) &  8 \\
      &           &          &          &        &           &         &            &    \\
\hline
\end{tabular}
\end{table}

\clearpage

\protect
\begin{table}
\caption{ 
{\bf Embeddings of the $SU(3)_c\otimes SU(2)_L$ states into the $\ol{\bf 27}$ in $E_6$. }
\vspace*{2ex}
Signs follow from our choice of projections, eqs.(\ref{eq:proj_SO10}-\ref{eq:proj_SU3xSU2}).
\vspace*{2ex}
}
\label{t:27barstates}
\tiny
\begin{tabular}{|c|rc|rc|rc|cr|}
\hline
                    &\multicolumn{2}{c|}{ } &\multicolumn{2}{c|}{ } &\multicolumn{2}{c|}{ } &\multicolumn{2}{c|}{ }\\
{\small Superfield} & 
                                 \multicolumn{2}{c|}{\small SU(3)$_c\otimes$SU(2)$_L$ } &
                                 \multicolumn{2}{c|}{\small SU(5)} &
                                 \multicolumn{2}{c|}{\small SO(10)} &
                                 \multicolumn{2}{c|}{\small E$_6$} \\
                    &\multicolumn{2}{c|}{ } &\multicolumn{2}{c|}{ } &\multicolumn{2}{c|}{ } 
                                                                                  &\multicolumn{2}{c|}{ } \\
standard &        &         &        &       &        &       &\multicolumn{2}{c|}{(100000) irrep}\\ 
embedding& weight &  irrep  & weight & irrep & weight & irrep & weight & level \\
                  &\multicolumn{2}{c|}{ } &\multicolumn{2}{c|}{ } &\multicolumn{2}{c|}{ } &\multicolumn{2}{c|}{ }\\
\hline
             &           &          &          &        &           &         &            &    \\
 $\ol{Q}$    &  (01)(1)  &  (01)(1) &  (0010)  & (0010) &  (00010)  & (00010) &  (000010)  &  0 \\
             &           &          &          &        &           &         &            &    \\
             & (1\1)(1)  &          &-(010\1)  &        & (1\1001)  &         & (01000\1)  &  4 \\
             &           &          &          &        &           &         &            &    \\
             & (\10)(1)  &          & (\101\1) &        &-(0\1010)  &         & (00001\1)  & 11 \\
             &           &          &          &        &           &         &            &    \\
             &  (01)(\1) &          & (1\101)  &        &-(0100\1)  &         & (\100001)  &  5 \\
             &           &          &          &        &           &         &            &    \\
             &-(1\1)(\1) &          &-(10\10)  &        & (100\10)  &         & (\1100\10) &  9 \\
             &           &          &          &        &           &         &            &    \\
             & (\10)(\1) &          & (0\100)  &        & (0000\1)  &         & (\100000)  & 16 \\
             &           &          &          &        &           &         &            &    \\
$\ol{U^c}$   &  (10)(0)  &  (10)(0) & (01\11)  &        & (001\10)  &         & (001\100)  &  2 \\
             &           &          &          &        &           &         &            &    \\
             & (\11)(0)  &          &-(\1001)  &        &-(\1010\1) &         & (0\11\110) &  9 \\
             &           &          &          &        &           &         &            &    \\
             &-(0\1)(0)  &          & (\11\10) &        &-(0\11\10) &         & (001\10\1) & 13 \\
             &           &          &          &        &           &         &            &    \\
$\ol{E^c}$   &  (00)(0)  &  (00)(0) & (1\11\1) &        &-(10\110)  &         & (\11\1100) &  7 \\
             &           &          &          &        &           &         &            &    \\
$\ol{D^c}$   &  (10)(0)  &  (10)(0) &  (1000)  & (1000) & (01\101)  &         & (01\1001)  &  3 \\
             &           &          &          &        &           &         &            &    \\
             & (\11)(0)  &          & (0\110)  &        &-(\11\110) &         & (00\1011)  & 10 \\
             &           &          &          &        &           &         &            &    \\
             &-(0\1)(0)  &          & (000\1)  &        &-(00\101)  &         & (01\1000)  & 14 \\
             &           &          &          &        &           &         &            &    \\
 $\ol{L}$    &  (00)(1)  &  (00)(1) & (\1100)  &        &-(\10001)  &         & (100\110)  &  7 \\
             &           &          &          &        &           &         &            &    \\
             &-(00)(\1)  &          & (00\11)  &        &-(\110\10) &         & (000\101)  & 12 \\
             &           &          &          &        &           &         &            &    \\
$\ol{N^c}$   & -(00)(0)  &  (00)(0) & -(0000)  & (0000) &-(1\110\1) &         & (\10100\1) &  6 \\
             &           &          &          &        &           &         &            &    \\
$\ol{T^c}$   &  (10)(0)  &  (10)(0) &  (1000)  & (1000) &  (10000)  & (10000) & (0001\10)  &  1 \\
             &           &          &          &        &           &         &            &    \\
             & (\11)(0)  &          & (0\110)  &        &-(0001\1)  &         & (0\10100)  &  8 \\
             &           &          &          &        &           &         &            &    \\
             &-(0\1)(0)  &          & (000\1)  &        & (1\1000)  &         & (0001\1\1) & 12 \\
             &           &          &          &        &           &         &            &    \\
$\ol{H_d}$   & -(00)(1)  &  (00)(1) &-(\1100)  &        &-(0\1100)  &         & (1\1100\1) &  5 \\
             &           &          &          &        &           &         &            &    \\
             & (00)(\1)  &          &-(00\11)  &        & (001\1\1) &         & (0\110\10) & 10 \\
             &           &          &          &        &           &         &            &    \\
 $\ol{T}$    & -(01)(0)  &  (01)(0) & -(0001)  & (0001) &-(\11000)  &         & (1\10001)  &  4 \\
             &           &          &          &        &           &         &            &    \\
             & (1\1)(0)  &          & (01\10)  &        & (000\11)  &         & (1000\10)  &  8 \\
             &           &          &          &        &           &         &            &    \\
             &-(\10)(0)  &          & (\1000)  &        &-(\10000)  &         & (1\10000)  & 15 \\
             &           &          &          &        &           &         &            &    \\
$\ol{H_u}$   & -(00)(1)  &  (00)(1) &-(001\1)  &        &-(00\111)  &         & (10\1100)  &  6 \\
             &           &          &          &        &           &         &            &    \\
             & (00)(\1)  &          & (1\100)  &        & (01\100)  &         & (00\11\11) & 11 \\
             &           &          &          &        &           &         &            &    \\
 $\ol{S}$    &  (00)(0)  &  (00)(0) &  (0000)  & (0000) &  (00000)  & (00000) & (\110\110) &  8 \\
             &           &          &          &        &           &         &            &    \\
\hline
\end{tabular}
\end{table}

\clearpage

\protect
\begin{table}[t]
\caption{ 
{\bf Embeddings of the $SU(3)_c\otimes SU(2)_L$ states into the non-zero weight states of the ${\bf 78}$ in $E_6$. }
Signs follow from our choice of projections, eqs.(\ref{eq:proj_SO10}-\ref{eq:proj_SU3xSU2}).
}
\label{t:78states}
\tiny
\begin{tabular}{|c|rc|rc|rc|cr|}
\hline
                    &\multicolumn{2}{c|}{ } &\multicolumn{2}{c|}{ } &\multicolumn{2}{c|}{ } &\multicolumn{2}{c|}{ }\\
{\small Superfield} & 
                                 \multicolumn{2}{c|}{\small SU(3)$_c\otimes$SU(2)$_L$ } &
                                 \multicolumn{2}{c|}{\small SU(5)} &
                                 \multicolumn{2}{c|}{\small SO(10)} &
                                 \multicolumn{2}{c|}{\small E$_6$} \\
                    &\multicolumn{2}{c|}{ } &\multicolumn{2}{c|}{ } &\multicolumn{2}{c|}{ } 
                                                                                  &\multicolumn{2}{c|}{ } \\
         &        &         &        &       &        &       &\multicolumn{2}{c|}{(100000) irrep}\\ 
         & weight &  irrep  & weight & irrep & weight & irrep & weight & level \\
                  &\multicolumn{2}{c|}{ } &\multicolumn{2}{c|}{ } &\multicolumn{2}{c|}{ } &\multicolumn{2}{c|}{ }\\
\hline
                &           &          &          &        &           &         &            &    \\
 $G$            &  (11)(0)  &  (11)(0) &  (1001)  & (1001) &  (01000)  & (01000) &  (000001)  &  0 \\
                &-(2\1)(0)  &          &-(11\10)  &        &-(100\11)  &         & (0100\10)  &  4 \\
                & (\12)(0)  &          & (0\111)  &        & (\1101\1) &         & (0\10011)  &  7 \\
                &-(1\2)(0)  &          & (01\1\1) &        &-(1\10\11) &         & (0100\1\1) & 15 \\
                &-(\21)(0)  &          & (\1\110) &        & (\1001\1) &         & (0\10010)  & 18 \\
                &-(\1\1)(0) &          &-(\100\1) &        & (0\1000)  &         & (00000\1)  & 22 \\
 $W$            &  (00)(2)  &  (00)(2) & (\111\1) &        & (0\1011)  &         & (10001\1)  &  6 \\
                & (00)(\2)  &          &-(1\1\11) &        & (010\1\1) &         & (\1000\11) & 16 \\
 $X$            &  (10)(1)  &  (10)(1) &  (101\1) &        &  (10\111) &         &  (01\1100) &  2 \\
                & (\11)(1)  &          & (0\12\1) &        & (00\120)  &         & (00\1110)  &  9 \\
                & (0\1)(1)  &          &-(001\2)  &        & (1\1\111) &         & (01\110\1) & 13 \\
                &  (10)(\1) &          & (2\100)  &        & (11\100)  &         & (\11\11\11)&  7 \\
                & (\11)(\1) &          & (1\210)  &        &-(01\11\1) &         & (\10\1101) & 14 \\
                & (0\1)(\1) &          &-(1\10\1) &        & (10\100)  &         & (\11\11\10)& 18 \\
$\ol{X}$        & -(01)(1)  &  (01)(1) & -(\1101) &        &  (\10100) &         & (1\11\110) &  4 \\
                & (1\1)(1)  &          & (\12\10) &        & (0\11\11) &         & (101\10\1) &  8 \\
                & (\10)(1)  &          &-(\2100)  &        & (\1\1100) &         & (1\11\11\1)& 15 \\
                & -(01)(\1) &          & (00\12)  &        & (\111\1\1)&         & (0\11\101) &  9 \\
                & (1\1)(\1) &          &-(01\21)  &        &-(001\20)  &         & (001\1\10) & 13 \\
                & (\10)(\1) &          & (\10\11) &        & (\101\1\1)&         & (0\11\100) & 20 \\
 $Q_{45}$       &  (10)(1)  &  (10)(1) &  (0100)  & (0100) & (1\1100)  &         & (00100\1)  &  1 \\
                & (\11)(1)  &          &-(\1010)  &        & (0\111\1) &         & (0\1101\1) &  8 \\
                & (0\1)(1)  &          & (\110\1) &        & (1\2100)  &         & (00100\2)  & 12 \\
                & (10)(\1)  &          &-(10\11)  &        & (101\1\1) &         & (\1010\10) &  6 \\
                & (\11)(\1) &          &-(0\101)  &        &-(0010\2)  &         & (\1\11000) & 13 \\
                & (0\1)(\1) &          &-(00\10)  &        & (1\11\1\1)&         & (\1010\1\1)& 17 \\
$U^c_{45}$      &  (01)(0)  &  (01)(0) & (1\110)  &        &-(1001\1)  &         & (\100100)  &  4 \\
                &-(1\1)(0)  &          & (100\1)  &        & (2\1000)  &         & (\1101\1\1)&  8 \\
                &-(\10)(0)  &          & (0\11\1) &        &-(1\101\1) &         & (\10010\1) & 15 \\
$E^c_{45}$      &-(00)(0)   &  (00)(0) &-(\11\11) &        & (0\12\1\1)&         & (0\12\10\1)& 10 \\
$\ol{Q}_{45}$   &  (01)(1)  &  (01)(1) & (0010)   & (0010) & (\11\111) &         & (10\1011)  &  5 \\
                &-(1\1)(1)  &          & (010\1)  &        & (00\102)  &         & (11\1000)  &  9 \\
                &-(\10)(1)  &          &-(\101\1) &        & (\10\111) &         & (10\1010)  & 16 \\
                &  (01)(\1) &          & (1\101)  &        & (\12\100) &         & (00\1002)  & 10 \\
                &-(1\1)(\1) &          &-(10\10)  &        &-(01\1\11) &         & (01\10\11) & 14 \\
                &-(\10)(\1) &          &-(0\100)  &        & (\11\100) &         & (00\1001)  & 21 \\
$\ol{U^c}_{45}$ &  (10)(0)  &  (10)(0) & (01\11)  &        & (\110\11) &         & (100\101)  &  7 \\
                & (\11)(0)  &          &-(\1001)  &        & (\21000)  &         & (1\10\111) & 14 \\
                & (0\1)(0)  &          &-(\11\10) &        & (\100\11) &         & (100\100)  & 18 \\
$\ol{E^c}_{45}$ &  (00)(0)  &  (00)(0) & (1\11\1) &        & (01\211)  &         & (01\2101)  & 12 \\
 $Q_{16}$       &  (10)(1)  &  (10)(1) &  (0100)  & (0100) &  (00001)  & (00001) & (010\110)  &  3 \\
                & (\11)(1)  &          &-(\1010)  &        & (\10010)  &         & (000\120)  & 10 \\
                & (0\1)(1)  &          & (\110\1) &        & (0\1001)  &         & (010\11\1) & 14 \\
                &  (10)(\1) &          &-(10\11)  &        &-(010\10)  &         & (\110\101) &  8 \\
                & (\11)(\1) &          &-(0\101)  &        &-(\1100\1) &         & (\100\111) & 15 \\
                & (0\1)(\1) &          &-(00\10)  &        &-(000\10)  &         & (\110\100) & 19 \\
$U^c_{16}$      &-(01)(0)   &  (01)(0) &-(1\110)  &        &-(01\110)  &         & (\11\1011) &  6 \\
                & (1\1)(0)  &          &-(100\1)  &        &-(10\101)  &         & (\12\1000) & 10 \\
                & (\10)(0)  &          &-(0\11\1) &        &-(00\110)  &         & (\11\1010) & 17 \\
$E^c_{16}$      &-(00)(0)   &  (00)(0) &-(\11\11) &        & (\101\10) &         & (001\210)  & 12 \\
$D^c_{16}$      &-(01)(0)   &  (01)(0) &-(0001)   & (0001) &-(0010\1)  &         & (\101\110) &  5 \\
                & (1\1)(0)  &          & (01\10)  &        &-(1\11\10) &         & (\111\10\1)&  9 \\
                & (\10)(0)  &          &-(\1000)  &        &-(0\110\1) &         & (\101\11\1)& 16 \\
$L_{16}$        & (00)(1)   &  (00)(1) & (001\1)  &        &-(1\1010)  &         & (\11001\1) &  7 \\
                & (00)(\1)  &          & (1\100)  &        & (1000\1)  &         & (\210000)  & 12 \\
$N^c_{16}$      & (00)(0)   &  (00)(0) & (0000)   & (0000) & (\11\101) &         & (01\1\111) & 13 \\
$\ol{Q}_{\ol{16}}$   &  (01)(1)  &  (01)(1) &  (0010)  & (0010) &  (00010)  & (00010) & (1\10100)  &  3 \\
                     &-(1\1)(1)  &          & (010\1)  &        &-(1\1001)  &         & (1001\1\1) &  7 \\
                     &-(\10)(1)  &          &-(\101\1) &        & (0\1010)  &         & (1\1010\1) & 14 \\
                     &  (01)(\1) &          & (1\101)  &        &-(0100\1)  &         & (0\101\11) &  8 \\
                     &-(1\1)(\1) &          &-(10\10)  &        & (100\10)  &         & (0001\20)  & 12 \\
                     &-(\10)(\1) &          &-(0\100)  &        &-(0000\1)  &         & (0\101\10) & 19 \\
$\ol{U^c}_{\ol{16}}$ &-(10)(0)   &  (10)(0) &-(01\11)  &        &-(001\10)  &         & (1\110\10) &  5 \\
                     &-(\11)(0)  &          & (\1001)  &        & (\1010\1) &         & (1\21000)  & 12 \\
                     &-(0\1)(0)  &          & (\11\10) &        &-(0\11\10) &         & (1\110\1\1)& 16 \\
$\ol{E^c}_{\ol{16}}$ & (00)(0)   &  (00)(0) & (1\11\1) &        &-(10\110)  &         & (00\12\10) & 10 \\
$\ol{D^c}_{\ol{16}}$ &-(10)(0)   &  (10)(0) &-(1000)   & (1000) &-(01\101)  &         & (10\11\11) &  6 \\
                     &-(\11)(0)  &          &-(0\110)  &        & (\11\110) &         & (1\1\1101) & 13 \\
                     &-(0\1)(0)  &          & (000\1)  &        &-(00\101)  &         & (10\11\10) & 17 \\
$\ol{L}_{\ol{16}}$   &-(00)(1)   &  (00)(1) &-(\1100)  &        & (\10001)  &         & (2\10000)  & 10 \\
                     &-(00)(\1)  &          & (00\11)  &        &-(\110\10) &         & (1\100\11) & 15 \\
$\ol{N^c}_{\ol{16}}$ &-(00)(0)   &  (00)(0) &-(0000)   & (0000) &-(1\110\1) &         & (0\111\1\1)&  9 \\
                     &           &          &          &        &           &         &            &    \\
\hline
\end{tabular}
\end{table}

\end{document}